\pdfoutput=1
\documentclass[fleqn,usenatbib]{mnras}

\usepackage{newtxtext,newtxmath}
\usepackage{orcidlink}

\usepackage[T1]{fontenc}

\DeclareRobustCommand{\VAN}[3]{#2}
\let\VANthebibliography\thebibliography
\def\thebibliography{\DeclareRobustCommand{\VAN}[3]{##3}\VANthebibliography}

\usepackage{graphicx}	
\usepackage{amsmath}	

\usepackage{booktabs}

\usepackage{slantsc}
\usepackage{ulem}
\usepackage{scalerel}
\newcommand\HI{H\protect\scaleto{$I$}{1.2ex}}
\usepackage{CJK}

\let\oldAA\AA
\renewcommand{\AA}{\text{\normalfont\oldAA}}

\begin{document}
\begin{CJK*}{UTF8}{gbsn}

\title[Massive Red Spiral Galaxies in MaNGA]{Massive Red Spiral Galaxies in SDSS-IV MaNGA Survey}

\author[J. Cui et al.]{
Jiantong Cui (崔健童)~\orcidlink{0009-0002-5336-5962},$^{1,2}$
Qiusheng Gu (顾秋生)~\orcidlink{0000-0002-3890-3729},$^{1,2}$\thanks{E-mail: qsgu@nju.edu.cn}
and Yong Shi (施勇)~\orcidlink{0000-0002-8614-6275}$^{1,2}$
\\
$^{1}$School of Astronomy and Space Science, Nanjing University, Nanjing 210093, P.R.China\\
$^{2}$Key Laboratory of Modern Astronomy and Astrophysics (Nanjing), Ministry of Education, Nanjing 210093, China\\
}

\date{Accepted 2024 January 10. Received 2023 December 26; in original form 2023 September 30}

\pubyear{2024}

\label{firstpage}
\pagerange{\pageref{firstpage}--\pageref{lastpage}}
\maketitle

\begin{abstract}
Massive red spiral galaxies (MRSGs) are supposed to be the possible progenitors of lenticular galaxies (S0s). We select a large sample of MRSGs ($M_*>10^{10.5}\rm M_{\odot}$) from MaNGA DR17 using the $g-r$ color vs. stellar mass diagram, along with control samples of blue spirals and S0s.
Our main results are as follows:
(1) After comparing the S$\rm \acute{e}$rsic index, concentration parameter, asymmetry parameter distribution, size-mass relation and $\Sigma_1$ (stellar mass surface density within the central 1 kpc)-mass relation, we find MRSGs are similar to S0s and have more compact and symmetric structures than blue spirals. MRSGs also resemble S0s in Dn4000, metallicity, Mgb/$\rm \left \langle Fe \right \rangle$ and $V/\sigma$ radial profile.
(2) By using MaNGA 2D spectra data, we separate the spatial regions into inner (R \textless 0.8$R_{\rm e}$) and outer (0.8$R_{\rm e}$ \textless R \textless 1.5$R_{\rm e}$) regions, and detect residual star formation in the outer regions of MRSGs.
(3) When we select a sub-sample of MRSGs with NUV$-r$ \textgreater 5, we find that they are completely star-formation quenched in both inner and outer regions. Compared to optically selected MRSGs, NUV$-r$ selected MRSGs appear to be more concentrated and have more massive dark matter halos.
The similarities between S0s and MRSGs suggest the possible evolutionary trend between MRSGs and S0s.

\end{abstract}




\section{Introduction}

The morphologies of galaxies contain the information of their formation and evolution. \cite{Hubble1936} classified galaxies into different types based on their appearances. S0 galaxies serve as a bridge connecting ellipticals and spirals in this morphological classification system. However, it is important to note that the morphological classification scheme does not directly indicate the evolutionary trajectory of galaxies.
In general, early-type massive galaxies are considered to form through major mergers between two spiral galaxies, leading to the rapid depletion of gas. On the other hand, late-type disc galaxies are believed to form within dark matter halos through a secular acquisition of gas from the surrounding environment \citep[e.g.][]{Fall&Efstathiou1980,Mo1998,Dutton2007,Munoz-Mateos2007}.
However, the formation pathways of S0s are believed to be diverse. Two commonly favored mechanisms are gas stripping via group infalls \citep[e.g.][]{Quilis2000,Crowl2005} and mergers \citep[e.g.][]{Tapia2017,Querejeta2015}.

On the other hand, galaxies exhibit a bimodal distribution on the color-mass diagram \citep[e.g.][]{Kauffmann2003b,Baldry2004,Baldry2006,Bell2004,Fabor2007,Ilbert2010}, SDSS photometric surveys have further revealed a close relationship between galaxy colors and morphologies \citep[e.g.][]{Strateva2001,Baldry2004,Schawinski2014}. Early-type galaxies tend to reside in the red sequence and late-type disc galaxies preferentially lie in the blue cloud. Many efforts have been devoted to investigate the mechanisms behind the transition of galaxies from blue to red sequence \citep[e.g.][]{Bell2004,Fabor2007,Marchesini2014,Schawinski2014}. The transition region between these two populations, known as the green valleys, occurs at different rates for different galaxies \citep{Schawinski2014}. Consequently, the star formation quenching is believed to be accompanied by a morphological transformation.

The discovery of red spirals challenged this scenario \citep{Dressler1999,Poggianti1999,Goto2003}. 
The red color observed in spirals can arise from either dust extinction in star-forming spirals or the passive spiral galaxies. In this work, we only focus on the truly quenched red spirals.
Red spirals are found in various environments. \cite{Masters2010} found red spirals are more likely in environment with intermediate densities, which can not fully explain their quenching.
One of the efficient ways to quench star formation is through bulge building \citep[e.g.][]{Martig2009,Bluck2014,Mendel2014}. \cite{Bundy2010} found red spirals have more concentrated light distribution compared with their blue counterparts. The fraction of red spirals with bars appears to be higher than normal spirals \citep[e.g.][]{Masters2010,Fraser-McKelvie2018,Guo2020}.

With the advancement of large photometric and spectroscopic galaxy surveys, such as SDSS \citep{York2000}, GALEX \citep{Martin2005}, WISE \citep{Wright2010}, more studies on the stellar population, structural properties and environment of red spirals have been carried out \citep[e.g.][]{Bundy2010,Masters2010,Robaina2012,Tojeiro2013,Fraser-McKelvie2018}.

Since cool gas is the raw material of star formation, the exhaustion of gas can be a major factor contributing to galaxy quenching. \cite{Guo2020} checked the \HI{} gas content of red spirals based on ALFAALFA data, and found red spirals have a comparable \HI{} detection rate as blue spirals but exhibit a lower ratio of \HI{} to stellar mass, thus the low star formation rate may be caused by the low gas surface density. \cite{Wang2022} extended this study by carrying out \HI{} observation to the rest of red spirals in \cite{Guo2020} not covered by ALFAALFA. 
They checked the color profile of red spirals in \cite{Guo2020} and found galaxies with \HI{} detection have bluer outer disks, indicating there are residual star formation in the outer regions.

The quenching mechanisms of red spirals are still under debate and show great complexity. On the other hand, some studies suggested the traditionally optically selected red spirals may be not completely quenched. \cite{Cortese2012} found that optically selected red spirals in \cite{Masters2010} still have star formation and UV emission. A large portion of the optically selected massive red spirals in \cite{Guo2020} fall on the green valley and blue cloud regions of NUV$-r$ color-mass diagram, as revealed by \cite{Zhou2021}, where they also found a non-negligible fraction of spaxels hold star formation in optically selected red spirals. Most of these spaxels are located in the low mass surface density regions, which are typical of the outer regions of galaxy disks.

In this work, we carry out detailed spectroscopic analysis to a large sample of massive red spirals using the two-dimensional spectrum obtained from the Mapping Nearby Galaxies at Apache Point Observatory \citep[MaNGA;][]{Bundy2015} and compare with blue spirals and S0s.

This paper is organized as follows. In Section~\ref{sec:data}, we introduce our sample selection and physical parameters. In Section~\ref{sec:results}, we carry out detailed analysis on the spectroscopic and structural properties of massive red spiral galaxies. 
We discuss the properties of NUV$-r$ selected red spirals and compare them with the characteristics of optically-selected red spirals in Section~\ref{sec:discussion}. We also explore the potential impact of environment and the evolutionary connection between red spirals and S0s. The main conclusions of our results are presented in Section~\ref{sec:conclusions}. Through the work, we adopt a Chabrier IMF \citep{Chabrier2003} and a $\Lambda$CMD cosmology with $\Omega_{\rm M}$$=0.3$, $\Omega_{\Lambda}$$=0.7$ and $\rm H_0$ $=$ 70 km$\cdot$s$^{-1}$$\cdot$Mpc$^{-1}$.

\section{Data}
\label{sec:data}

\subsection{Sample Selection}
MaNGA is one of the core projects of SDSS-IV \citep{Blanton2017,Bundy2015}, which aims at mapping detailed components and structures of nearby galaxies using integral field unit (IFU) spectroscopy.
For our study, we select samples from the 16th data release of MaNGA \citep{Abdurrouf2022ApJS} and adopt the morphology classification from MaNGA Morphology Deep Learning DR17 catalog \citep{Dominguez2022}, using a criteria as follows:

Spirals: T$-$Type \textgreater 0 

S0s: T$-$Type $\leq$ 0 and P\_S0 \textgreater 0.5

Ellipticals: T$-$Type $\leq$ 0 and P\_S0 $\leq$ 0.5

The stellar mass, star formation rate and $g-r$ color information are taken from MPA-JHU DR7 \citep{Kauffmann2003,Brinchmann2004} and NYU Value-Added Galaxy Catalog \citep{Blanton2005}, respectively. The Petrosian magnitudes are dust-corrected and applied k-correction to z$=$0. 
Following \cite{Speagle2014ApJS}, we take the IMF conversion as below.
\begin{equation}
M_{*, K}=1.06 M_{*, C}=0.62 M_{*, S}
\end{equation}
where the subscripts represents Kroupa, Chabrier and Salpeter IMFs respectively.
As the MPA-JHU catalog adopts a Kroupa initial mass function (IMF) \citep{Kroupa2001}, we subtract 0.025 dex to convert the stellar mass and SFR to the Chabrier IMF \citep{Chabrier2003}. There are 9262 galaxies in total, consisting of 2312 ellipticals, 874 S0s and 6076 spirals.

We subsequently apply a mass cut to $M_*\geq10^{10.5}$ M$_{\odot}$, which yields 1619 spirals, 637 S0s and 1849 ellipticals. In order to reduce the impact of internal dust extinction, we rule out S0s and spirals with b/a \textless 0.5 to avoid the edge-on effect. After this step, we have 400 S0s and 863 spirals.

\begin{figure}
\centering
\includegraphics[scale=1]{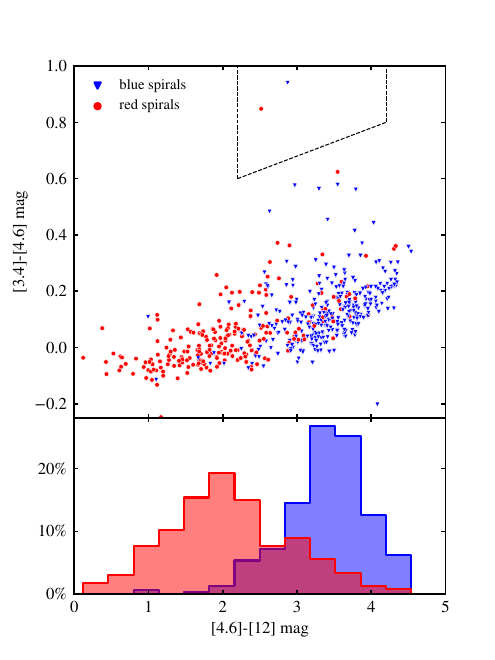}
\caption{The WISE color-color distribution of red spirals and blue spirals. The top panel shows the distribution of red spirals and blue spirals on the WISE \citep{Wright2010} color-color diagram. The AGN region is defined by \protect\cite{Jarrett2011ApJ}, shown as the dashed line. The bottom panel shows the histogram of [4.6]-[12] color for red spirals and blue spirals separately. Most of the dusty blue spirals lie in the region where [4.6]-[12]>2.5 mag. We adopt this value to distinguish dusty MRSGs from truly quenched MRSGs.}
\label{wise}
\end{figure}

\begin{figure}`  
\centering
\includegraphics[scale=1]{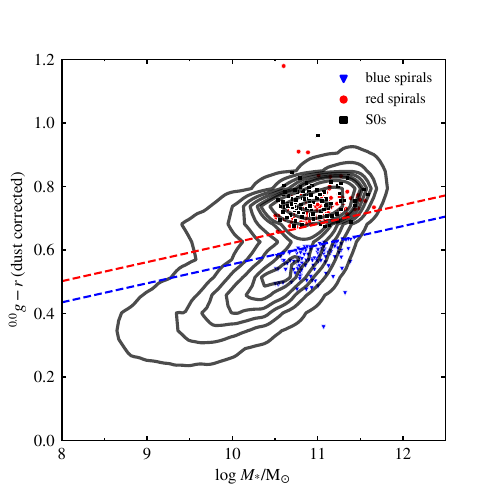}
\caption{The g-r color-mass diagram of our selected samples. The dashed lines show the separation between red sequence, green valley and blue cloud as defined in Equation \ref{equ:colormass1} and \ref{equ:colormass2}. The green valley is defined as a strip which spans one-third of the color difference between the centroid of red sequence and blue cloud \citep{Chen2010}. The slope is derived from the best-fit of the galaxy number peak at red sequence. MRSGs, blue spirals and S0s are shown as red, blue and black dots. The contours show the number density distribution of SDSS DR7 \citep{York2000} galaxies.}
\label{color_mass}
\end{figure}

To select massive red spirals (MRSGs), we make use of the $g-r$ color-mass diagram \citep[e.g.][]{Bluck2014,Mendel2014} as shown in Fig \ref{color_mass}. We fit the red sequence as a straight line at the point where the number of galaxies peaks, and the slope is consistent with the result of \cite{KT2016}. The green valley is defined as a trip which spans one-third of the color difference between the centroid of red sequence and blue cloud \citep{Chen2010}. The value is roughly equal to the mean sigma value obtained from gaussian fitting in each mass bin of the red sequence's galaxy count.
The boundary of green valley is defined as below
\begin{equation}
{ }^{0.0} g-{ }^{0.0} r=0.06 \log \left(M / \mathrm{M}_{\odot}\right)+0.023 \label{equ:colormass1}
\end{equation}
\begin{equation}
{ }^{0.0} g-{ }^{0.0} r=0.06 \log \left(M / \mathrm{M}_{\odot}\right)-0.043 \label{equ:colormass2}
\end{equation}
where $^{0.0} g$ and $^{0.0} r$ are $g$-band and $r$-band magnitudes K-corrected to z=0.
Among them, 254 spirals are located in the red sequence, while 341 spirals are in the blue cloud. However, it is important to note that the color of some MRSGs may be attributed to dust absorption rather than the absence of on-going star formation. To account for this, we utilize WISE \citep{Wright2010} color to exclude dusty MRSGs \citep{Mahajan2020}, as shown in Fig \ref{wise}. Noticing most of the blue spiral galaxies reside in the region where [4.6]-[12]>2.5 mag, we apply a color cut of [4.6]-[12]=2.5 to distinguish quenched MRSGs from their dusty counterparts. We also exclude galaxies which fall in the AGN region defined in \cite{Jarrett2011ApJ}, marked by the dashed lines in Fig \ref{wise}. During this step, we remove 61 dusty spirals and have 169 MRSGs.

To select control samples, we only use galaxies which have bulge and total mass information from \cite{Mendel2014} and obtain their bulge mass and total mass for each galaxy separately. The bulge-to-total mass ratio is defined as B/T $=$ $M_{\rm bulge}$/($M_{\rm bulge}$+$M_{\rm disk}$). After this step, we have 131 MRSGs, 367 S0s and 265 blue spirals.

Although star formation in most S0s has ceased, recent studies show that some S0 galaxies still exhibit on-going star formation, suggesting they may have experienced different formation pathways compared to quenched S0s \citep[e.g.][]{Xu2022,Rathore2022}.
To ensure a clear comparison, we choose to focus only on quenched S0s as our control samples. The star-forming S0s are excluded according to the boundary between green valley and passive galaxies defined in \cite{Trussler2020}, which yields 296 normal S0s.

\begin{figure*}
\centering
\includegraphics[scale=1]{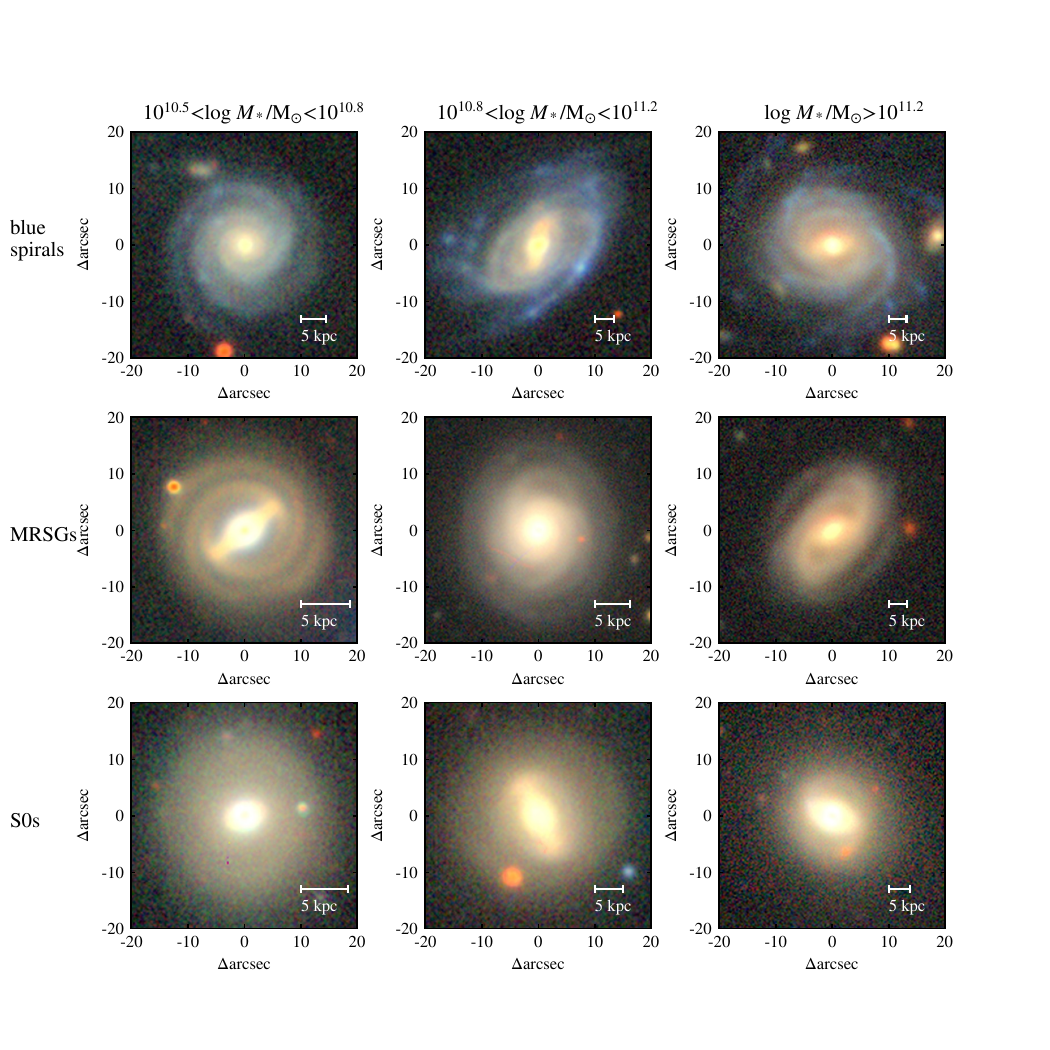}
\caption{DESI false-color ($g$-,$r$-,and $z$-bands) images \citep{Dey2019AJ} of MRSGs, blue spirals and S0s in different mass ranges. The top row shows images of blue spiral samples in mass range $\rm 10^{10.5}M_{\odot}$-$\rm 10^{10.8}M_{\odot}$, $\rm 10^{10.8}M_{\odot}$-$\rm 10^{11.2}M_{\odot}$ and $M_*$>$\rm 10^{11.2}M_{\odot}$, respectively. The middle row and bottom row displays the images of MRSGs and S0s in the same order. A scale bar representing a physical scale of 5 kpc is shown in the lower right corner of each panel.}
\label{image}
\end{figure*}

We match each MRSG with a control galaxy from the selected S0 galaxies and blue spirals with the criteria $\Delta$$M_*$ \textless 0.15, $\Delta$B/T \textless 0.05, and $\Delta$z \textless 0.03. This yields 112 S0s and 117 blue spirals as control samples. We combine these with the 131 MRSGs, resulting in a total of 360 galaxies as our final sample. The distribution of these galaxies on the color-mass diagram is shown in Fig \ref{color_mass}.
Examples of DESI false-color ($g$-,$r$-,and $z$-bands) images \citep{Dey2019AJ} of MRSGs and control samples in different mass ranges are shown in Fig \ref{image}.

\subsection{Physical Parameters}

The spatially resolved properties such as Dn4000, Mgb/$\rm \left \langle Fe \right \rangle$ and EW(H$\alpha$), as well as the radial velocity and velocity dispersion are taken from the MaNGA Data Analysis Pipeline \citep[DAP;][]{Belfiore2019,Westfall2019}. We correct redshift for radial velocity to ensure the central velocity is zero, and correct the velocity dispersion for instrumental effects and beam swearing \citep{Westfall2019}. To convert radial velocity to rotation velocity, we follow the method of \cite{Deeley2020}.

The inclination is estimated from the ellipticity, where $\epsilon$ is the ellipticity taken from the DAP and q$_0$ is taken to be 0.2 \citep{Tully2000}.
\begin{equation}
\cos i=\sqrt{\frac{(1-\epsilon)^2-q_0^2}{1-q_0^2}}
\end{equation}
Then the rotational velocity is recovered following the equation below:
\begin{equation}
v_t=\frac{v_r}{\sin i \cos \theta}
\end{equation}
where $v_t$ and $v_r$ are the rotational velocity and radial velocity respectively. The azimuthal angle $\theta$ is taken from the DAP file. Only spaxels within 45 degrees from the major axis are included to avoid divergence.

The spatially resolved star formation rate is estimated using a conversion factor of H$\alpha$ luminosity from \cite{Kennicutt1998}:
\begin{equation}
\operatorname{SFR}\left(\mathrm{M}_{\odot} \mathrm{yr}^{-1}\right)=7.9 \times 10^{-42} L(\mathrm{H} \alpha)[\rm erg\cdot s^{-1}] \label{equ:sfr}
\end{equation}
The H$\alpha$ flux is taken from MaNGA Pipe3d catalog and corrected for attenuation using the equation below:
\begin{equation}
A(\mathrm{H} \alpha)=\frac{K_{\mathrm{H} \alpha}}{-0.4\left(K_{\mathrm{H} \alpha}-K_{\mathrm{H} \beta}\right)} \times \log \left(\frac{F_{\mathrm{H} \alpha} / F_{\mathrm{H} \beta}}{2.86}\right)
\end{equation}
where $\kappa_{\mathrm{H}\alpha}$ and $\kappa _{\mathrm{H}\beta}$ denote the value of extinction curve evaluated at the wavelength of $\mathrm{H}\alpha$ and $\mathrm{H}\beta$.
\begin{equation}
f_{X, \text { corr }}=f_{X, \text { obs }} \times 10^{A(\mathrm{H} \alpha) / 2.5}
\end{equation}

We obtain the effective radius, metallicity and emission line information of our galaxies from the MaNGA Pipe3D value added catalog for DR17 \citep{Sanchez2016}. We utilize this catalog to calculate $\Sigma_1$, which is defined as the mass density within 1 kpc distance from the galaxy center. Since they adopts a Salpeter IMF, We subtract 0.24 dex mass offset to match the Chabrier IMF.

The MaNGA Data Reduction Pipeline \citep[DRP;][]{Law2016} processes the raw data and provides flux-calibrated, sky-subtracted, regularly-gridded data cubes for each individual exposure of a given galaxy. In our study, we use files that have a logarithmic-based wavelength solution for the spectrum stacking process.

We get the structural parameters such as concentration index and asymmetry index from \cite{Meert2015}, where they perform photometric decompositions for galaxies in SDSS DR7. We get the projected number densities for galaxies from GEMA-VAC \citep{Argudo2015,Etherington2015,Wang2016}.

The flux-weighted stellar kinematics are also studied using the equation below \citep{Cappellari2007}.
\begin{equation}
\left(\frac{V}{\sigma}\right)_{\mathrm{e}}^2 \equiv \frac{\left\langle V^2\right\rangle}{\left\langle\sigma^2\right\rangle}=\frac{\sum_{n=1}^N F_n V_n^2}{\sum_{n=1}^N F_n \sigma_n^2} \label{equ:v_sigma}
\end{equation}
where $F_n$, $V_n$ and $\sigma_n$ are the g-band flux, line-of-sight velocity and velocity dispersion of each sapxel respectively.

To study the gas content of our samples, we derive \HI{} detection information from \HI{}-MaNGA \citep{Masters2019}, a program that follows up on MaNGA galaxies using observations from ALFAALFA \citep{Haynes2018} and the Green Bank Telescope (GBT). 
We get the halo mass for our samples from \cite{Yang2007}. They provide two sets of masses calculated based on characteristic stellar mass and characteristic luminosity respectively, and we choose the former as the halo masses for our galaxies. 

\section{Results}
\label{sec:results}

\subsection{Radial Distribution of Physical Properties}

\begin{figure}
\centering
\includegraphics[scale=1]{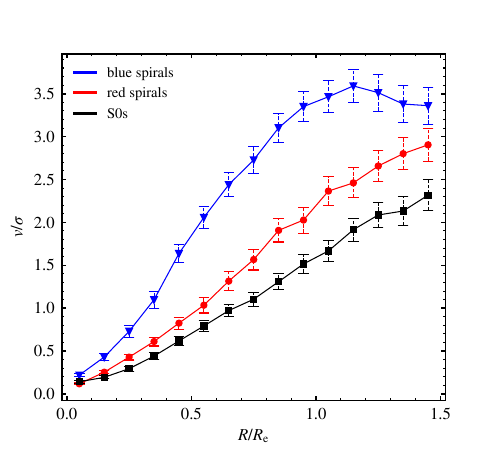}
\caption{The median radial profile of the ratio of rotational velocity to velocity dispersion for MRSGs and the control samples, with a radial bin of 0.1$R_e$. The error bar represents the error of the median.}
\label{v_sigma}
\end{figure}

We check the radial distribution of basic physical properties of galaxies.
First, we examine the kinematics of our samples to see if they show profiles consistent with a system holding a bulge and a disk.
The radial profiles of rotational velocity to velocity dispersion ratios are illustrated in Figure \ref{v_sigma}. All galaxies exhibit a dispersion-dominated bulge and a rotation-supported disk, which are consistent with their optical morphologies. However, the disks of MRSGs and S0 galaxies tend to be more dispersion-dominated. In the outer disk, the V/$\sigma$ of MRSGs is $\sim$1.5 \sout{dex} lower than that of blue spirals, indicating the spiral arms of MRSGs may be fading away.

\begin{figure*}
\centering
\includegraphics[scale=1]{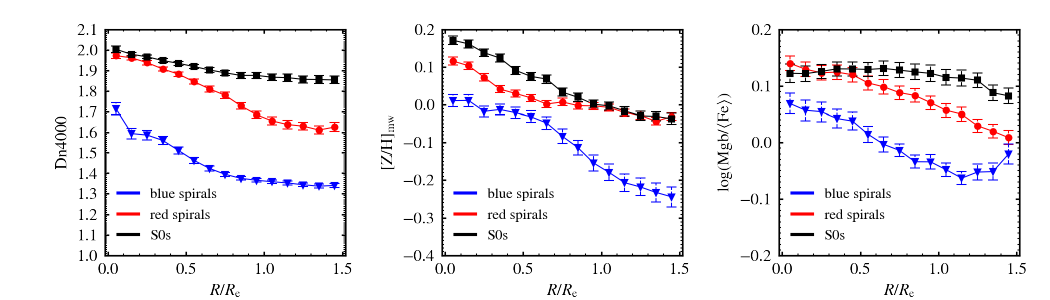}
\caption{The median radial profile of Dn4000, metallicity and Mgb/$\rm \left \langle Fe \right \rangle$ for MRSGs and the control samples. The value of each galaxy in each radial bin is determined as the median value of the spaxels within that bin. The median radial profile is obtained by calculating the median value across all galaxies of each type. The red line represents MRSGs, blue line represents blue spirals and black line represents S0 galaxies. The error bars show the error of the median.}
\label{global}
\end{figure*}

The left panel of Figure \ref{global} presents the median Dn4000 values for each radial bin of our sample galaxies. Dn4000 is defined as the flux ratio between 4050$\AA$$-$4100$\AA$ and 3090$\AA$$-$3095$\AA$, which is a good tracer of galaxy stellar age and is sensitive to metallicity \citep[e.g.][]{Kauffmann2003b}.
Both MRSGs and S0s exhibit similar old bulges, but the profile of MRSGs is steeper. The stellar populations in the outer disks of MRSGs are younger.
Blue spirals, on the other hand, have central regions with steeper gradient and flat disks. The median Dn4000 value for blue spirals is $\sim$0.4 lower than that of MRSGs, indicating a younger stellar population with an extended star formation history.

The spatial metallicity distributions are presented in the middle panel of Figure \ref{global}. The metallicities and errors are taken from Pip3d catalog, they are derived from spectrum fitting using a Monte-Carlo procedure. The typical error from fitting procedure is $\sim$0.5 dex for each spaxel \citep{Sanchez2016}. The steep metallicity gradient is expected from a monolithic-like collapse in the deep potential well of galaxies, in which case the gas can be retained efficiently and lead to longer star formation \citep{Tortora2010}. The low value and steep slope again suggest the extended star formation of blue spirals. 
MRSGs exhibit high metallicity with shallow profiles.
S0 galaxies also show high metallicity. They are more metal-rich than MRSGs in the inner regions but become more metal poor in the outer regions, which is similar to the behavior of elliptical galaxies as reported in \cite{Hao2019}. Mergers or AGN feedback can smear out the stellar populations or quench the galaxies on global scale, which shallows the metallicity gradients \citep[e.g.][]{White1980,Kobayashi2004,Tortora2009}.

The Mgb/$\rm \left \langle Fe \right \rangle$ ratio is defined as Mgb/$\rm \left \langle Fe \right \rangle$=Mgb/0.5(Fe5270+Fe5335). It serves as a proxy for $\alpha$/Fe abundance, which traces the relative strength of intense starburst compared to long-term star formation \citep[e.g.][]{Matteucci1986,Thomas2005}. The value of Mgb/$\rm \left \langle Fe \right \rangle$ can reflect the timescale of galaxy formation. The longer star formation timescale is accompanied by lower Mgb/$\rm \left \langle Fe \right \rangle$ ratio \citep{Thomas2005}.
As shown in the right panel of Fig \ref{global}, both MRSGs and S0 galaxies exhibit similarly high Mgb/$\rm \left \langle Fe \right \rangle$ ratios in their central regions, denoting a fast formation of bulge. The identical short timescales indicates the similar formation mechanisms. However, the Mgb/$\rm \left \langle Fe \right \rangle$ value decreases in the outer regions of MRSGs, which may be caused by the residual star formation. 
Blue spirals have a steep slope. The bulge of blue spirals has formed within a rather short timescale, while the disk undergoes a more extended star formation process. This is consistent with an inside-out growth scenario \citep[e.g.][]{Mo1998,Mateos2007}.
The different formation timescales indicate MRSGs have different formation mechanisms compared to nearby blue spirals. They should not be regarded simply as faded blue star-forming spirals.

We notice that our conclusions are consistent with that of \cite{Hao2019}, although the inner and outer profiles of MRSGs in our study are not as shallow as those in \cite{Hao2019}. This difference may be attributed to the larger sample size in our study. Our sample includes a greater number of MRSGs at different stages of evolution, resulting in a steeper gradient.

\subsection{Spectroscopic Properties}
\label{sec:spectroscopic}

\subsubsection{Spatially Resolved Star Formation}
\label{sec:spatialsfr}
Previous works have revealed that the optically selected MRSGs are not totally quenched\citep[e.g.][]{Cortese2012,Zhou2021}. 
In order to check the location and magnitude of these residual star formation as well as the differences between outer and inner regions of massive MRSGs, we use the two-dimension galaxy spectrum obtained from MaNGA Data Reduction Pipeline \citep[DRP;][]{Law2016} to quantatively measure these properties.

\begin{figure*}
\centering
\includegraphics[scale=1]{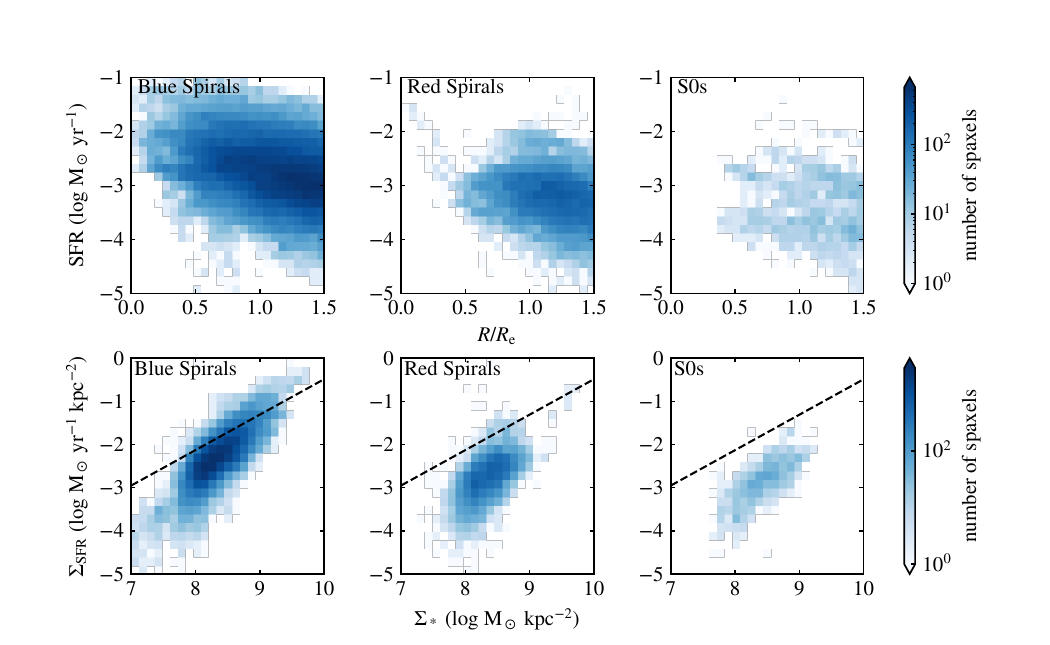}
\caption{The distribution of spatailly resolved star formation rate for star-forming spaxels in MRSGs and the control samples. The colorbar represent the number of spaxels. The left, middle and right panel represents the distribution of blue spirals, MRSGs and S0s respectively. The bottom row shows the spatially resolved star formation relation for spaxels in our samples, where the dashed lines show the main sequence from \protect\cite{Enia2020} as comparison. Star-forming spaxels in MRSGs are below the main sequence.}
\label{spatial_sfr}
\end{figure*}

To check where the star formation locates within our MRSG samples, we get the emission line flux and equivalent width from the MaNGA Pipe3d catalog \citep{Sanchez2016}. We identify spaxels with EW(H$\alpha$) \textgreater 3 and log(\ion{N}{ii}/H$\alpha$) \textless -0.4 as regions undergoing star formation \citep{Fernandes2011}. The star formation rates are estimated using a conversion factor of H$\alpha$ luminosity following Equation \ref{equ:sfr}. The distributions of these star-forming spaxels are shown in Fig \ref{spatial_sfr}, with the color bar indicating the number of spaxels.

The fraction of the star-forming spaxels in MRSGs (16\%) is lower compared to their blue counterparts (53\%). However, it is still significantly higher than the fraction of star-forming spaxels in S0 galaxies (5\%). 

The average star formation rate of star-forming spaxels in MRSGs is also roughly 1 dex lower than that in blue spirals. Although residual star formation still exists in MRSGs, it falls below the resolved star-forming main sequence as revealed by \cite{Zhou2021}. 

From Fig \ref{spatial_sfr} we can see that the distribution of star-forming spaxels in blue spirals spans the entire galaxy, while MRSGs are concentrated in the outer regions where $R/R_{\rm e}$>0.8. We define regions inside a 0.8 $R_{\rm e}$ aperture as inner regions and out of this as outer regions. In the following part, we divide the galaxies into inner ($R$<0.8$R_{\rm e}$) and outer (0.8$R_{\rm e}$<$R$<1.5$R_{\rm e}$) regions to highlight the characteristics of star formation and other properties more clearly. The star formation rate in the star-forming zones of S0 galaxies is similar to that of MRSGs, but the distribution seems to be more spatially irregular.

\begin{figure*}
\centering
\includegraphics[scale=1.]{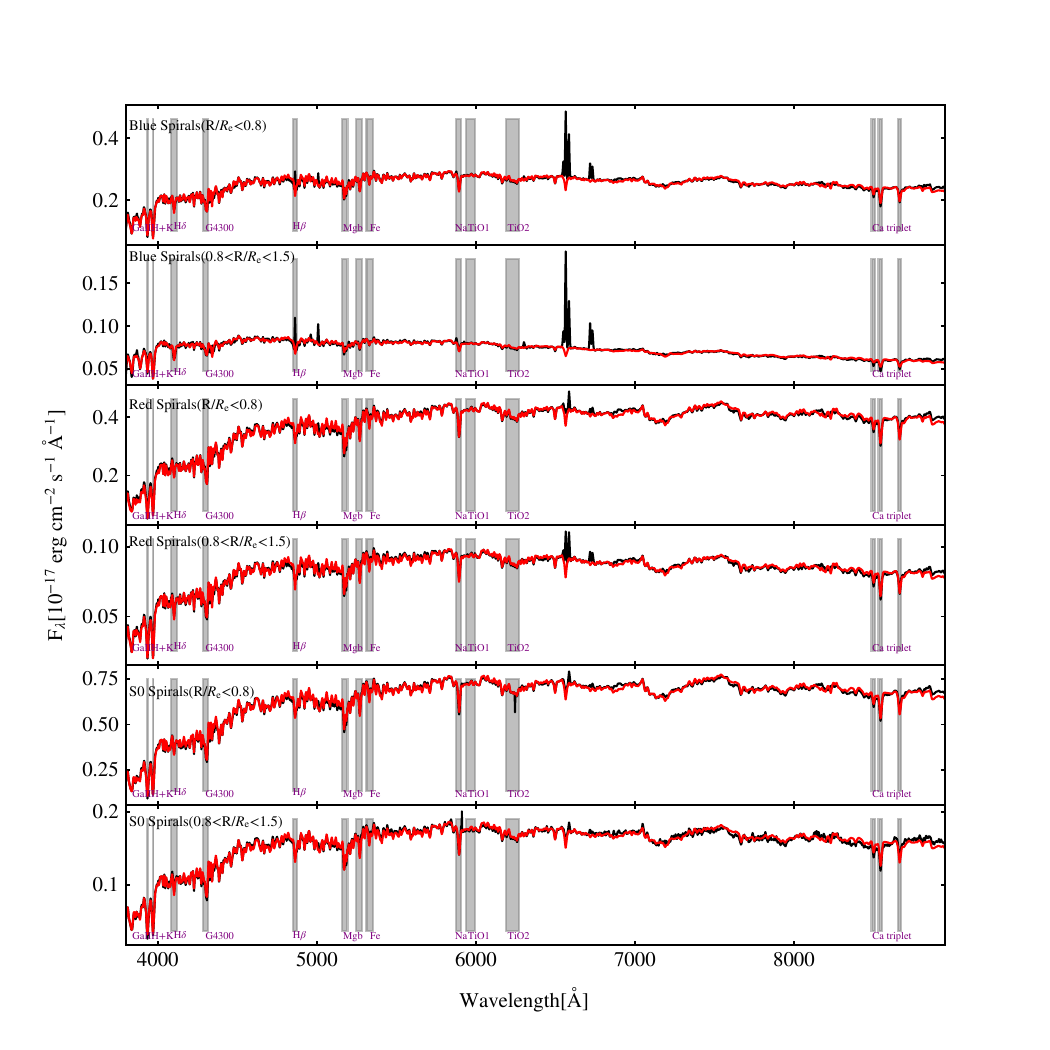}
\caption{The stacked spectra for outer and inner regions of MRSGs and comparison with control samples. The stacked spectra are shown as black lines and the best-fitting models are shown as the red lines. The spectrum profiles of MRSGs show characteristics of old stellar population like the ones of S0s. However, there is a clear H$\alpha$ emission in the outer part of MRSGs. Main absorption features are denoted by the shaded strip on the spectrum.}
\label{spec_fit}
\end{figure*}

\subsubsection{Different Behavior of Star Formation in Inner and Outer Regions}
In order to detect the possible residual star formation, we stack the spectrum within 0.8 $R_{\rm e}$ and out of 0.8 $R_{\rm e}$ separately following the method below. In order to eliminate the potential influence of AGN on the H$\alpha$ emission, we have excluded spaxels which fall within the AGN region in the WHAN diagram with EW(H$\alpha$) \textgreater 3 and log(\ion{N}{ii}/H$\alpha$) \textgreater -0.4 \citep{Fernandes2011}, which enables us to focus solely on the star formation properties of the galaxies.

First we deredshift the spectrum to the rest frame according to the equation below:

\begin{equation}
\lambda_{\mathrm{rest}}=\frac{\lambda_{\mathrm{obs}}}{(1+z)\left(1+\frac{v}{c}\right)}
\end{equation}
where z is the NSA redshift and v is the radial velocity of each spaxel retrieved from the DAP, then we explore linear interpolation to resample both the spectrum and error spectrum into a wavelength grid spanning from 3800$\AA$ to 9200$\AA$ with a sampling interval of 1$\AA$. 
We normalize each spectrum according to its continuum flux value at 5500$\AA$. 
After the de-redshift, resample, and continuum-normalization process, we stack the spectra within each type and location bin using the weighted average method. 
\begin{equation}
\left\langle f_{\mathrm{\nu}}\right\rangle=\frac{\sum_{i=1}^n\left(w_i \mathrm{f}_{\nu, i}\right)}{\sum_{i=1}^n w_i}
\end{equation}
where the weighted factor w$_i$=$\sigma_{i}^{-2}$ is the inverse variance of each spectrum at a given wavelength. The signal-to-noise ratio (S/N) of the co-added spectrum is improved significantly, which enables us to conduct a detailed comparison with stellar population synthesis models. It also allows for a clear observation of the H$\alpha$ emission in the outer regions of MRSGs. We use Voronoi-binned spectrum for the stacking process. The original S/N for individual spaxel in the center of galaxy is $\sim$50 and fall to $\sim$10 at 1.5 Re. After the stacking process, the S/N for the stacked spectrum reach to $\sim$100 times higher. The S/N of the stacked spectrum in Fig \ref{spec_fit} is $\sim$ 4000, 3000, 7000, 4000, 7000, 4000 respectively. During the stacking process, we did not consider the covariances between different pixels. Based on the factor which rescales the noise computed without accounting for covariance to those that do in \cite{Law2016}, the noise may be underestimated by a factor of $\sim$5. However, since the resulting (S/N) of the stacked spectrum is very high, this has only a minimal impact on the final results.

As shown in Fig \ref{spec_fit}, the spectral profiles of MRSGs and S0s are very similar, predominantly characterized by the presence of old stellar population. 
With the exception of the outer region of MRSGs, where a distinct H$\alpha$ emission, accompanied by weak [\ion{N}{ii}] and [\ion{S}{ii}] emissions, becomes evident, the spectra of both MRSGs and S0s exhibit minimal emission lines.

To furthur investigate the properties of the stellar populations in our samples, we fit each stacked spectrum with a set of model spectra using \textsc{starlight} \citep{CidFernandes2005}. The fitting procedure involves employing a combination of 138 model spectra, which compromises 6 metallicities and 23 different ages from BC03 \citep{Bruzual2003}. For this analysis, we employ models using the Padova 1994 evolutionary tracks and the Chabrier IMF.

We measure the H$\alpha$ flux in the outer regions of MRSGs and blue spirals using a gaussian fitting procedure and estimate the corresponding star formation rate. The measurement has been repeated 1000 times, each time we perturb the spectra with the corresponding error spectra. The standard deviation is taken as the measurement error.
The measured specific star formation rate per spaxel for the outer disk of MRSG in logarithmic scale is -11.33$\pm$0.08 yr$^{-1}$, and -10.51$\pm$0.02 yr$^{-1}$ and -10.97$\pm$0.02 yr$^{-1}$ for the outer and inner part of blue spirals respectively. 
The mass-weighted age of the inner and outer part of MRSGs are 12.17 Gyr and 10.75 Gyr respectively, while the mass-weighted age of blue spirals are 10.18 Gyr and 8.30 Gyr. The residual star formation in the disk of red spirals make a negative age gradient. These results are consistent with the resolved conclusion in Sec \ref{sec:spatialsfr}. 


\begin{table*}
    \centering
    \caption{The measurement of Lick indices of stacked spectrum}
    \label{tab1}
    \begin{tabular}{lcccrrr}
    \toprule
        \textbf{Index} & \textbf{blue inner} & \textbf{blue outer} & \textbf{red inner} & \textbf{red outer} & \textbf{S0 inner} & \textbf{S0 outer} \\ \midrule
        Dn4000 & 1.564  $\pm$ 0.001 & 1.356  $\pm$ 0.001 & 1.994 $\pm$ 0.001&1.698 $\pm$ 0.001&  1.971 $\pm$ 0.001& 1.837 $\pm$ 0.001\\ 
        H$\delta_{\rm A}$ &1.593 $\pm$ 0.003& 3.389 $\pm$ 0.003& -1.179 $\pm$ 0.003& 1.388 $\pm$ 0.004& -0.775 $\pm$ 0.003& 0.448 $\pm$ 0.004\\
        H$\delta_{\rm F}$ &1.556 $\pm$ 0.002& 2.184 $\pm$ 0.002& 0.284 $\pm$ 0.002& 1.480 $\pm$ 0.003& 0.512 $\pm$ 0.002& 1.051 $\pm$ 0.003\\
        CN$_1$ &-0.035 $\pm$ 0.001&-0.066 $\pm$ 0.001& 0.021 $\pm$ 0.001& -0.029 $\pm$ 0.001& 0.017 $\pm$ 0.001&-0.009 $\pm$ 0.001\\
        CN$_2$ & -0.000 $\pm$ 0.001& -0.029 $\pm$ 0.001& 0.053 $\pm$ 0.001& 0.008 $\pm$ 0.001& 0.051 $\pm$ 0.001& 0.026 $\pm$ 0.001\\
        Ca4227 &0.860 $\pm$ 0.002& 0.517 $\pm$ 0.002& 1.271 $\pm$ 0.001&1.000 $\pm$ 0.002&  1.338 $\pm$ 0.001& 1.146 $\pm$ 0.002\\
        G4300 &3.721 $\pm$ 0.003& 2.273 $\pm$ 0.003& 5.469 $\pm$ 0.002&4.103  $\pm$ 0.003& 5.196 $\pm$ 0.002& 4.715 $\pm$ 0.004\\
        H$\gamma_{\rm A}$ &-2.754 $\pm$ 0.003&-0.020 $\pm$ 0.003& -6.364 $\pm$ 0.002& -3.230 $\pm$ 0.004&-5.880 $\pm$ 0.003& -4.526 $\pm$ 0.004\\
        H$\gamma_{\rm F}$ &0.225 $\pm$ 0.002&1.462 $\pm$ 0.002& -1.394 $\pm$ 0.001& 0.134 $\pm$ 0.002& -1.122 $\pm$ 0.002&-0.449 $\pm$ 0.002\\
        Fe4383 & 4.175 $\pm$ 0.004& 2.707 $\pm$ 0.004& 5.783 $\pm$ 0.003&4.621 $\pm$ 0.004&  5.809 $\pm$ 0.003& 5.243 $\pm$ 0.005\\
        Fe4531 &2.525 $\pm$ 0.003& 1.745 $\pm$ 0.003& 3.277 $\pm$ 0.002& 2.800 $\pm$ 0.003& 3.305 $\pm$ 0.002& 3.062 $\pm$ 0.003\\
        C$_2$4668 & 4.132 $\pm$ 0.004&2.670 $\pm$ 0.005& 5.753 $\pm$ 0.003& 4.653 $\pm$ 0.005&  6.005 $\pm$ 0.003& 5.376 $\pm$ 0.005\\
        H$\beta$ & 2.146 $\pm$ 0.002&2.495 $\pm$ 0.002& 1.767 $\pm$ 0.001& 2.261 $\pm$ 0.002&  1.858 $\pm$ 0.001&2.116 $\pm$ 0.002\\
        Mg$_1$ &0.054 $\pm$ 0.001&0.031 $\pm$ 0.001& 0.070 $\pm$ 0.001& 0.058 $\pm$ 0.001& 0.075 $\pm$ 0.001& 0.064 $\pm$ 0.001\\
        Mg$_2$ &0.158 $\pm$ 0.001&0.104 $\pm$ 0.001& 0.202 $\pm$ 0.001& 0.171 $\pm$ 0.001& 0.212 $\pm$ 0.001& 0.189 $\pm$ 0.001\\
        Mgb &2.680 $\pm$ 0.002& 1.860 $\pm$ 0.002& 3.338 $\pm$ 0.001&2.847 $\pm$ 0.002& 3.452 $\pm$ 0.001& 3.144 $\pm$ 0.002\\
        Fe5270 & 2.380 $\pm$ 0.002&1.727 $\pm$ 0.003& 2.943 $\pm$ 0.001& 2.604 $\pm$ 0.002&  3.029 $\pm$ 0.002& 2.83 $\pm$ 0.003\\
        Fe5335 &2.143 $\pm$ 0.003&1.591 $\pm$ 0.003&  2.621 $\pm$ 0.002& 2.324 $\pm$ 0.003& 2.718 $\pm$ 0.002& 2.516 $\pm$ 0.003\\
        Na5895 & 2.517 $\pm$ 0.002& 1.953 $\pm$ 0.003& 2.951 $\pm$ 0.001& 2.683 $\pm$ 0.002&  3.073 $\pm$ 0.001& 2.824 $\pm$ 0.002\\
        TiO$_1$ &0.033 $\pm$ 0.001&0.027 $\pm$ 0.001&  0.041 $\pm$ 0.001& 0.037 $\pm$ 0.001&  0.045 $\pm$ 0.001&0.038 $\pm$ 0.001\\
        TiO$_2$ &0.055 $\pm$ 0.001&0.040 $\pm$ 0.001&  0.071 $\pm$ 0.001& 0.062 $\pm$ 0.001&  0.080 $\pm$ 0.001&0.066 $\pm$ 0.001\\
        CaT &7.072 $\pm$ 0.004&6.224 $\pm$ 0.005&  7.615 $\pm$ 0.002&7.330 $\pm$ 0.004& 7.700 $\pm$ 0.002&7.533 $\pm$ 0.004\\
        CaT$_1$ &1.060 $\pm$ 0.002& 0.877 $\pm$ 0.002& 1.077 $\pm$ 0.001&1.085 $\pm$ 0.002& 1.084 $\pm$ 0.001& 1.086 $\pm$ 0.002\\
        CaT$_2$ & 3.056 $\pm$ 0.002&2.598 $\pm$ 0.003& 3.383 $\pm$ 0.001& 3.190 $\pm$ 0.002& 3.424 $\pm$ 0.001&3.307 $\pm$ 0.002\\
        CaT$_3$ & 2.971 $\pm$ 0.002& 2.668 $\pm$ 0.002& 3.159 $\pm$ 0.001& 3.067 $\pm$ 0.002& 3.179 $\pm$ 0.001& 3.132 $\pm$ 0.002\\
    \bottomrule
    \end{tabular}
\end{table*}


We have marked some typical absorption features on Fig \ref{spec_fit}. Noticing that there are a lot of absorption lines in the stacked spectra of MRSGs and S0 galaxies, we manage to quantatively describe these absoprtion features.
Lick indices are defined using two pseudo continuum bandpasses on either side of the index bandpass. We measure the absorption line features and the corresponding uncertainties from the stacked spectrum using \textsc{indexf} \citep{Cardiel2010} following the definition of \citep{Worthey1997,Trager1998}. The error is calculated by propagating random errors using the error spectrum \citep{Cardiel2010,Cardiel1998}.Gaussian noise determined by the input error spectrum is introduced in each pixel. Subsequently, index errors are obtained as the unbiased standard deviation of approximately 1000 measurements for each index. Given the high S/N in our stacked spectra, the uncertainties calculated in this process are minimal.

Since the spectral resolution of the best-fit of our stacked spectrum is 3 $\AA$, which is smaller than the resolution of Lick system, we convolve the fitting result with the original Lick resolution. The value of indices are corrected for broadening effects caused by the velocity dispersion using the correction curve described in \cite{Carson2010,Vazdekis2003}.
The final results are listed in Table \ref{tab1}.

It is obvious that MRSGs show higher Dn4000 than blue spirals, and the outer and inner regions of MRSGs have a larger difference than S0 galaxies do. Other indices such as \ion{Ca}{ii} H+K lines and \ion{Ca}{ii} triplet which are stronger in late-type stars are also more prominent in MRSGs, which is consistent with their older stellar age.

The Balmer absorption index can reflect the star formation timescale. MRSGs have lower H$\delta_{\rm A}$ value than blue spirals, which indicates they lack recent extended star formation. We can also find this trend in a series of Mg and Fe absorption lines.

MRSGs show more prominent TiO, Fe, Mg, \ion{Ca}{ii} H and K lines and are also more enriched in CN$_1$, CN$_2$, Ca4227 and G4300 absorptions.

In order to further understand the quenching scenario of MRSGs as indicated by their spectroscopic features, we discuss their structural properties and gas content in the following parts.

\subsection{Structural Properties}
\label{sec:structure}


\cite{VanderWel2014} claimed that galaxies with different star formation activities follow different size-mass relations. The slope of the relation remains almost constant at different redshifts with a changing zero point. They fit the size-mass relation with a power-law form $r\left(m_*\right) / \mathrm{kpc}=\mathrm{A} m_*^\alpha$, where $m_* \equiv M_* / 7 \times 10^{10} \mathrm{M}_{\odot}$. The $\alpha$ parameter for late-type galaxies is 0.25 and 0.75 for early-type galaxies. The selection of late-type and early-type galaxies are based on star formation activities. In our analysis, we use the size-mass relation at z=0.25 obtained from their fit for comparison.

\begin{figure}
\includegraphics[scale=1]{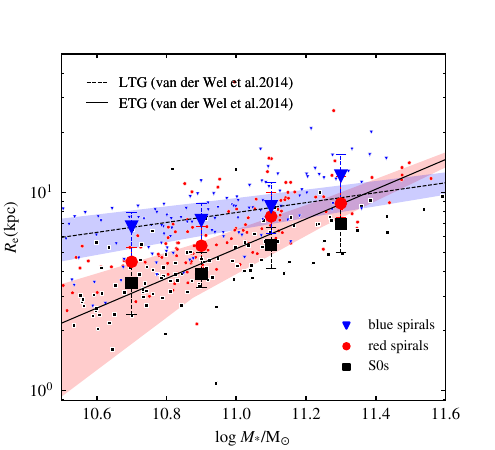}
\caption{The size-mass relation of MRSGs. The masses are divided into bins ranging from 10.6 to 11.4 with a bin width of 0.2 dex. The larger markers with error bars represent the median value and the corresponding 16th to 84th percentile range within each bin. The dashed and solid lines are from \protect\cite{VanderWel2014}, which are empirical relation for late-type and early-type galaxies respectively. The sizes of MRSGs are between those of blue spirals and S0 galaxies.}
\label{size_mass_fig}
\end{figure}


As shown in Fig \ref{size_mass_fig}, the sizes of MRSGs at a given mass bin are between blue spirals and S0s, suggesting MRSGs are more compact than normal blue spirals. The gas fractions in the progenitor disks of low-mass early-type galaxies are higher. This results in a stronger dissipative process, leading to a smaller size of low-mass early-type galaxies and a steeper gradient in the observed size-mass relation for early-type galaxies \citep{Hopkins2009ApJ}. MRSGs show a similar steep gradient as S0s, which may be explained by a similar dissipative scenario such as mergers.

We further examine the S$\rm \acute{e}$rsic index of MRSGs and the two control samples. The values of S$\rm \acute{e}$rsic index are taken from the S$\rm \acute{e}$rsic fits to the 2D surface brightness profiles of the final MaNGA DR17 galaxy sample from MaNGA PyMorph DR17 photometric catalogue \citep{Dominguez2022}. Almost half of the blue spirals have a S$\rm \acute{e}$rsic index below the critical value of n$=$2.5, indicating the prescence of pseudo-bulges. In contrast, both MRSGs and S0s exhibit larger S$\rm \acute{e}$rsic index values. 
The distributions of S$\rm \acute{e}$rsic index for MRSGs and blue spirals are different at a significance level >99.99\%, with p-value 1$\times 10^{-13}$. While MRSGs have simialr S$\rm \acute{e}$rsic index distributions with S0s, and the p-value is 0.16.

\begin{figure*}
\includegraphics[scale=1]{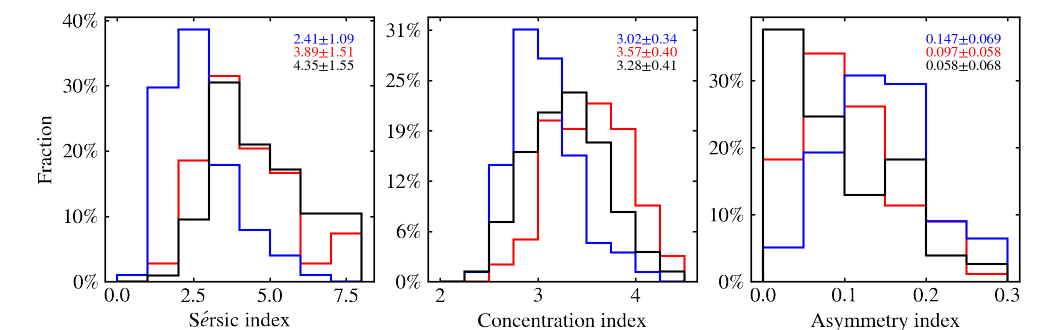}
\caption{The parameter distributions for MRSGs and the control samples. The red, blue, and black lines represent the distribution of MRSGs, blue spirals and S0s respectively. The median values and standard deviations are labeled in the upper right corner of each figure.}
\label{para_distribution}
\end{figure*}

We get the concentration index and asymmetry from \cite{Vikram2010}. The definations are as follows.
The concentration index is defined as the ratio of galaxy radius which contains 80 percent of total light to the radius which contains 20 percent of the total light. 
\begin{equation}
C=5 \log \left(\frac{r_{80}}{r_{20}}\right)
\end{equation}

The asymmetry index is defined as 
\begin{equation}
A_{\mathrm{O}}=\frac{\sum\left|I_{\mathrm{O}}-I_{\mathrm{R}}\right|}{\sum I_{\mathrm{O}}}
\end{equation}
where I$_O$ and I$_R$ are the flux of the original galaxy image and image rotated by 180 degrees about its center.

The distribution of concentration index and asymmetry index is shown in Fig \ref{para_distribution}. The mean measuremnet error of concentration index is 0.18 and 0.36 for asymmetry. Taking these into account, the error of mean of concentration index is $\sim$0.02, the mean value of MRSGs (3.60) is larger than that of blue spirals (3.07) at a great significance level. And the mean asymmetry value of MRSGs (0.09) is lower than that of blue spirals (0.14) at a significance level greater than 1$\sigma$, where the error of median of asymmetry is $\sim$0.03. The lower asymmetry index of MRSGs compared to blue spirals indicates the spiral structure in MRSGs is not as prominent as blue spirals.
Based on the distribution of S$\rm \acute{e}$rsic index and concentration index, we can deduce that the relatively small sizes of MRSGs are resulted from a concentrated structure and build-up process of bulge. 

\begin{figure}
\includegraphics[scale=1]{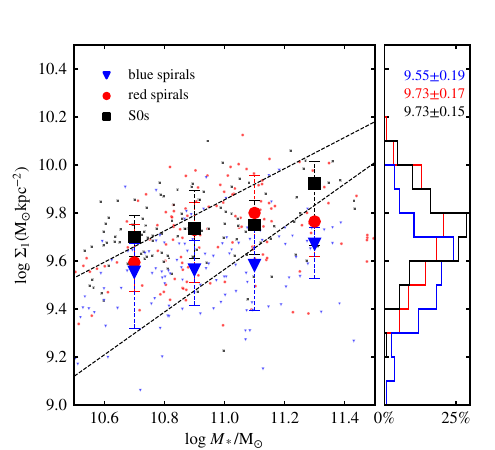}
\caption{The distribution of the central 1 kpc stellar mass surface density for our galaxy samples. The upper and lower dashed line are the empirical relation observed in quiescent and star-forming galaxies respectively. The median value and 16th to 84th percentile range distribution of sample galaxies are marked in each mass bin.}
\label{sigma_mass}
\end{figure}

Recent studies have suggested mass surface density of the core reigion as a tight indicator of galaxy quenching \citep[e.g.][]{Fang2013,Barro2017,Whitaker2017ApJ,Cheung2012,Chen2020ApJ}. Some critical thresholds of galaxy surface densities are found to be correlated with the quenching process. A fixed surface density threshold is proposed to trace galaxies quenching \cite[e.g.][]{Kauffmann2006,Franx2008}. \cite{Fang2013} recommended a changing critical value of surface density relative to stellar mass and found galaxies in red sequence and green valley follow a power-law relation of $\Sigma_1$ (stellar mass surface density within the central 1 kpc) and stellar mass. \cite{Barro2017} explored the scaling relations between the core mass density and stellar mass separately for both star-forming and quiescent galaxies, which we used in Fig \ref{sigma_mass} for comparison. We calculate the stellar mass surface density within the central 1 kpc by directly adding the mass of each spaxel within 1 kpc aperture derived from Pipe3d value-added catalog and divide the area. The distribution of $\Sigma_1$ and mass are shown in Figure \ref{sigma_mass}.
The distribution of $\Sigma_1$ of MRSGs peaks at $\rm 10^{9.73} M_{\odot}kpc^{-2}$, which is similar to S0 galaxies. Notice we matched the B/T when selecting blue spirals and S0s as our control samples, the distribution of their core density is not representative of the whole S0 and blue spiral populations, but the distribution still shows significant difference. Most of the MRSGs have reached a characteristic density for early-type galaxies at their center.


The scaling relation observed in early-type galaxies requires formation of concentrated structure via a gas-rich phase, such as mergers or contractions \citep{VanderWel2014}. \cite{VanderWel2014} found a population of late-type galaxies with small sizes that span the entire size range of early-type galaxies, and these galaxies maintain nearly constant number densites between z=3 and z=1.5. \cite{Barro2017} also discovered a population of compact star-forming galaxies across a wide redshift range from z=3 and z=0.5. These galaxies may undergo contraction and serve as the progenitors of early-type galaxies.
The presence of these galaxies suggests the existence of strong dissipative processes which move the gas towards central regions and form a dense core. Mergers or compaction due to disk instability may be the possible mechanisms \citep{Dekel2014MNRAS,Shen2003}. 
After experiencing the gas-rich dissipative phase at high redshift, the star formation rapidly quenched in these compact galaxies. They gradually grow their stellar mass along the observed scaling relation through minor mergers or accretions subsequently \citep{Barro2017}.

MRSGs have compact sizes and follow the observed trend for early-type galaxies. Additionally, they hold high Mgb/<Fe> values, which suggests they experience similar strong dissipative process during bulge formation. Simulations also show gas-rich major mergers at higher redshift can produce galaxies with disk morphologies\citep[e.g.][]{Deeley2021,Springel2005,Robertson2006,Hopkins2009,Athanassoula2016,Sparre2017}. 
However, the observed scaling relation itself has large scatters mainly due to the relatively strength of dissipative and dissipationless process in different galaxies. On the other hand, the threshold of galaxy structural parameter is found to be a necessary but not sufficient condition for galaxies to be quenched \citep{Cheung2012,Bell2012}. These can account for the scatters of the distribution of MRSGs.

Although the size of our samples is relatively small, which weakens the significance of the results, the trend is still evident. We see a clear similarity of core density distribution between MRSGs and S0 galaxies. The trend as well as the S$\rm \acute{e}$rsic index and concentration index together indicates the concentrated structures of MRSGs may result from a rapid bulge growth.

\section{Discussion}
\label{sec:discussion}

\subsection{NUV-r Selected MRSGs}
Since traditional optically selected MRSGs are not totally quenched, some studies suggested using NUV$-r$ color to identify weak star formation activities \citep[e.g.][]{Cortese2012,Zhou2021}. \cite{Zhou2021} found NUV$-r$ selected MRSGs hold high Dn4000 while the optically selected MRSGs have younger stellar population in the disk. In order to verify this from the spectrum, we separate NUV$-r$ selected ones from the whole MRSGs samples and stack their spectra in the outer regions.

We obtain the ultraviolet measurements for our samples from the NASA-Sloan Atlas catalog \citep[NSA;][]{Blanton2005} and select 21 MRSGs with NUV$-r$>5 identified as NUV$-r$ selected MRSGs. 

\begin{figure*}
\includegraphics[scale=1]{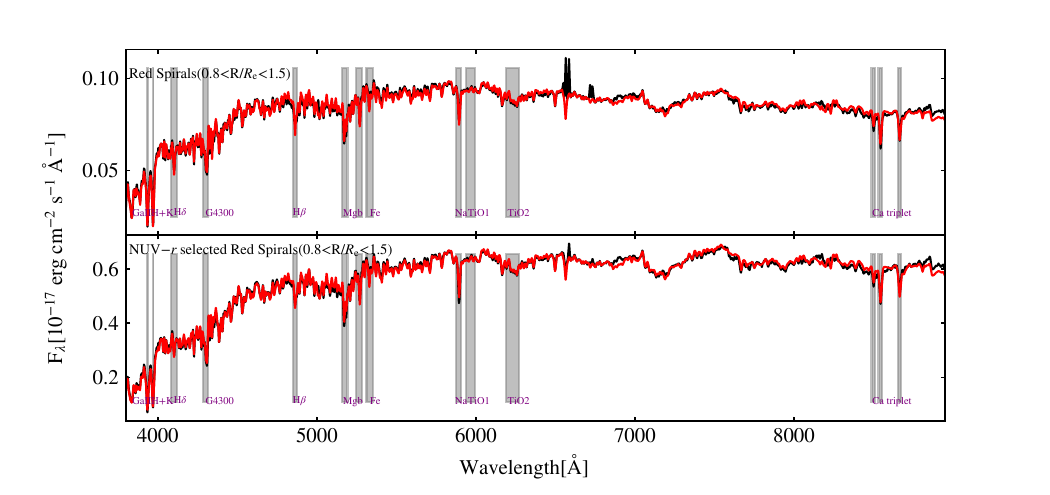}
\caption{The stacked spectrum for the outer regions of MRSGs and NUV$-r$ selected MRSGs. The stacked spectra are shown as black lines and the best-fitting models are shown as the red lines. There is no obvious H$\alpha$ emission in the NUV$-r$ selected MRSGs.}
\label{spec_uv}
\end{figure*}

As shown in Fig \ref{spec_uv}, the stacked spectra of the outer regions of NUV$-r$ selected MRSGs and the whole set of MRSGs have similar profiles, but they show clearly different H$\alpha$ emission features. No obvious H$\alpha$ emission can be seen from the NUV$-r$ selected MRSGs, indicating the lack of residual star formation on the disk. Result of stellar population fitting shows the disks of NUV$-r$ selected MRSGs are older than the remaining red ones, with a mass-weighted age equals 11.6 Gyr.

To figure out whether NUV$-r$ selected MRSGs hold different morphologies, we check the structural parameters for NUV$-r$ selected MRSGs. The NUV$-r$ selected MRSGs follow the empirical size-mass relation observed in early-type galaxies more tightly. We apply K-S test by comparing the distribution of NUV$-r$ selected MRSGs and simulated random points following empirical relation for early-type and late-type galaxies separately. The resulting p-values are 0.79 and 0.01. The whole MRSGs population have similar gradient while average larger sizes than early-type galaxies. The level of similarity between NUV$-r$ selected MRSGs and early-type galaxies is much higher.

We use similar method to generate random points along the empirical $\Sigma_1$-mass relation for quiescent and star-forming galaxies respectively and apply K-S test between them and NUV$-r$ selected MRSGs. The p-values are 0.43 and 0.001 respectively, suggesting greater similarities between NUV$-r$ selected MRSGs and quiescent galaxies. This is consistent with the size-mass relation for NUV$-r$ selected MRSGs. 

\begin{figure}
\centering
\includegraphics[scale=1]{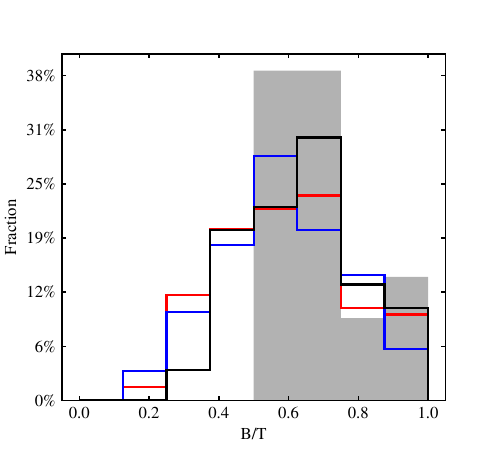}
\caption{The B/T distribution for our galaxy samples. The B/T ratio of MRSGs, blue spirals and S0s have been matched, thus having a similar distribution. The distribution for NUV$-r$ selected MRSGs is represented by the gray shadow, which concentrates in the region where B/T>0.5.}
\label{bt_distribution}
\end{figure}

The distribution of B/T of NUV-r selected MRSGs are plotted in Fig \ref{bt_distribution}. Since we have matched B/T at the sample selection procedure, the B/T distributions of our samples are similar. However, while optically selected MRSGs have a rather uniform B/T distribution, NUV$-r$ selected MRSGs only have B/T ratios larger than 0.5, indicating a bulge-dominant morphology.

Besides the dissipative process mentioned in Section \ref{sec:structure}, bars may also play a role in the growth of bulges. The bar may facilitate the gas redistribution from the discs of galaxies to the centers and provide ingredients for BH accretion and starburst in the past \citep[e.g.][]{CS1981,Shlosman2000ApJ}. The bar fraction of NUV$-r$ selected MRSGs (69.90\%) is slightly higher than that of MRSGs (56.48\%) and blue spirals (49.57\%), However, the $\chi^2$ test shows no significant difference, with p-value equals 0.82 and 0.42 respectively. The relative higher fraction of bars is also observed in \cite{Masters2010}. Considering the small size of our samples, statistics using larger samples are needed to confirm the conclusions. However, this secular evolution alone can not explain the quick formation of compact centers in MRSGs.

\begin{figure}
\centering
\includegraphics{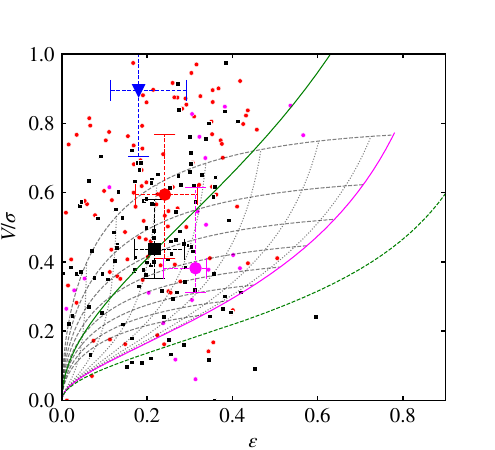}
\caption{($V/\sigma$,$\epsilon$) diagram for our galaxy samples, where $V/\sigma$ is defined in \ref{equ:v_sigma} and $\epsilon$ is the ellipticity. The background lines are defined following \protect\cite{Cappellari2007}. The green line represents the relation for edge-on isotropic rotators. The magenta line corresponds to $\delta$=0.7$\varepsilon_{intr}$. The green dashed line is 1/3 the isotropic line, which is a rough separation for regular rotators and non-regular rotators. The gray dotted line is the mangenta line derived with different inclination angle. And the gray dashed line is for different intrinsic ellipcity. The median value and the corresponding 16th to 84th percentile range within each bin distribution of MRSGs, NUV$-r$ selected MRSGs, blue spirals and S0s are shown as red, magenta, blue and black stars with error bars on the plot.}
\label{kinematics_plot}
\end{figure}

The distributions of galaxy samples on the ($V/\sigma$,$\epsilon$) diagram are shown in Fig \ref{kinematics_plot}. Since we have excluded galaxies with b/a>0.5 to avoid the edge-on effect, the samples concentrate on the left side of the panel.
We mark the median values of galaxies' $V/\sigma$ on the plot. 
We adopt the K-S test for MRSGs and blue spirals, and the p-value is 2$\times 10^{-9}$. The result shows that the distribution of $V/\sigma$ for red spirals is different from blue spirals with a significance level > 99.99\%. The decreasing of ratio between ordered rotation and random motions with increasing concentration was shown in previous study \citep[e.g.][]{Fogarty2015,Krajnovic2013}.
The median $V/\sigma$ for NUV$-r$ selected MRSGs is smaller than MRSGs, and is even slightly smaller than that of S0 galaxies, which suggests the random motions occupy a larger proportion in the kinematics of NUV$-r$ selected MRSGs.
For S0s and MRSGs, the p-value of K-S test is 4$\times 10^{-5}$, while the p-value is 0.11 for S0s and NUV-r selected MRSGs. The result shows NUV-r selected MRSGs show more similar distribution with S0s than MRSGs.
Most of MRSGs and S0s are encompassed by the envelope of the magenta line and its projections, where regular rotators tend to locate. We also notice that several MRSGs and S0s fall below the green dashed line, which is an approximate separation between regular rotators and non-regular rotators. The relatively low fraction is consistent with previous works \citep[e.g.][]{Deeley2020}. 

Compared to optically selected MRSGs, NUV$-r$ selected MRSGs have more compact structures and the random motion gradually dominate the kinematics.  

\begin{figure}
\centering
\includegraphics[scale=1]{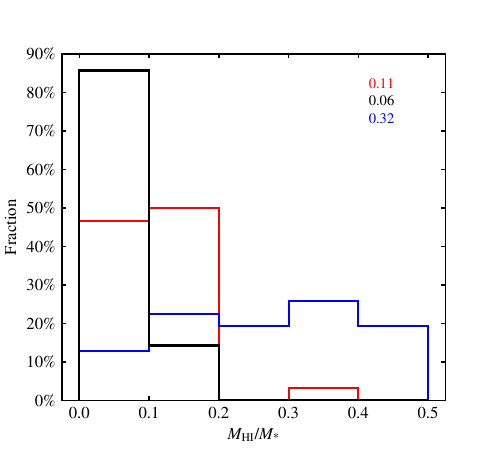}
\caption{The ratio of HI gas mass over stellar mass for our galaxy samples. MRSGs, blue spirals and S0s are represented by red, blue and black lines respectively.The median values are labeled in the upper right corner.}
\label{HI_mass}
\end{figure}

The quenching of star formation is ultimately related to the the process of removing the ingredient gas or preventing it from cooling and collapsing. 
We get the gas information from the combination of \HI{} MaNGA program \citep{Masters2019,Stark2021}, ALFAALFA survey \citep{Giovanelli2005,Haynes2018} and FAST observation carried out by \cite{Wang2022}. 
64 MRSGs, 37 blue spirals and 45 S0 galaxies in our samples are included in the above surveys. The \HI{} detection rate of MRSGs is 46.88\%, which is consistent with the findings in \cite{Guo2020}. Blue spirals and S0s have 89.19\% and 15.56\% \HI{} detection rate respectively, which coheres with their strength of star formation activities. The \HI{} detection rate of NUV$-r$ selected MRSGs drops to 25\% compared to the whole MRSG samples. The \HI{} mass to stellar mass ratio of MRSGs is also lower than that of blue spirals, as revealed by Fig \ref{HI_mass}.
The remaining gas contents in MRSGs provide the ingredient for the residual star formation. As revealed by the work of \cite{Wang2022}, these gas may exit in the disks of MRSGs. Our work is consistent with this point of view as the residual H$\alpha$ emission remains in the outer regions of MRSGs. On the other hand, the remaining gas may not transform to star formation effectively because of their low surface density as analyzed in \cite{Guo2020}.
In the future, spatially resolved \HI{} and molecular gas detection can be carried out to verify the scenario and the star formation law for these galaxies.

The above results show NUV$-r$ selected MRSGs lack residual star formation, and their structures are more compact than optically-selected MRSGs. The use of NUV$-r$ color can be a more accurate criterion for selecting totally quenched MRSGs. The different spectroscopic and structural properties of NUV$-r$ selected MRSGs indicate they are closer to total star formation quenching.

\subsection{Environment Effects}

\cite{Schawinski2014} found environment plays a vital role in the quenching of late-type galaxies in their study, where late-type galaxies in massive halos (>10$^{12}$ M$_\odot$) concentrate in the red sequence while galaxies with less massive halos assemble in the blue cloud. The result holds the same for MRSGs.

\begin{figure*}
\centering
\includegraphics[scale=1]{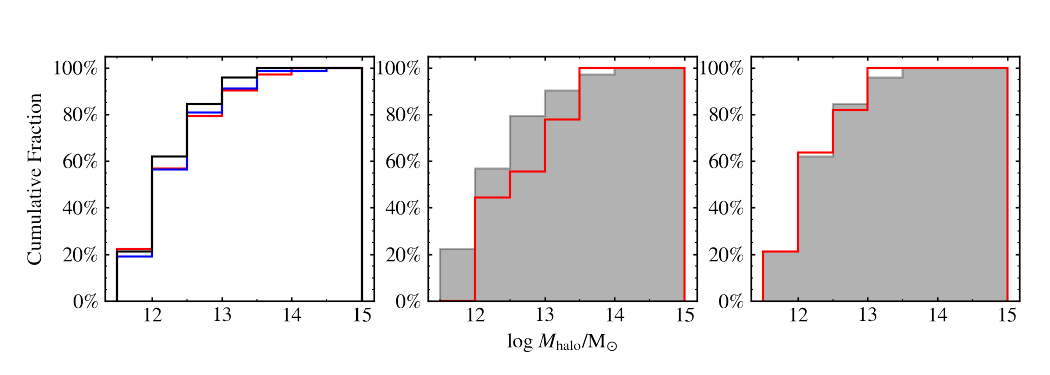}
\caption{The left panel shows the dark matter halo mass cumulative fraction for our galaxy samples at different stellar mass. MRSGs, blue spirals and S0s are shown in red, blue and black color. The middle panel shows the distribution for NUV$-r$ selected MRSGs in red lines, and the gray shadow shows the distribution for the whole MRSGs samples for comparison. The right panel shows the cumulative halo mass fraction for S0 galaxies and S0 galaxies with NUV$-r$>5.}
\label{halo}
\end{figure*}

There are 75.98\% (98/129) MRSGs, 84.35\% (97/115) blue spirals and 77.07\% (84/109) S0s central galaxies (labelled as brightest in the catalog of \cite{Yang2007}).
The cumulative fractions of halo masses for the central galaxies in our samples are shown in Fig \ref{halo}. 
The distribution of halo masses for MRSGs and NUV$-r$ selected MRSGs have a clear difference. While about 80\% of MRSGs have dark halo masses exceeding the critical halo mass of $10^{12}\rm M_{\odot}$ for halo quenching, all of the NUV$-r$ selected MRSGs are above this value, and their halo masses tend to concentrate at the high mass end. 

We also plot the distribution for S0s and S0s with NUV$-r$>5 for comparison, there isn't clear difference between these two groups of galaxies.

The different halo mass distributions between MRSGs and NUV$-r$ selected MRSGs are consistent with the fact that NUV$-r$ selected MRSGs have older stellar population. \cite{Zhou2021} found NUV$-r$ selected MRSGs ceased their star formation earlier through a Bayesian analysis of star foramtion history. 
The shock heating in massive halos is efficient to prevent fresh gas inflows and also make the gas easier to be heated and pushed by the central black holes (BHs) after the halo mass reached a critical value \citep{Dekel2006}.
Since these NUV$-r$ selected MRSGs have more compact structures than the whole MRSGs population, they tend to hold BHs with higher masses \citep[e.g.][]{Fang2013}, which may contribute to their star formation quenching due to more cumulative feedback. The growing halo furthur in turn contribute to their star formation quenching.
After the quenching of bulge promoted by the halo quenching and possible BH feedback, the change of morphology may furthur stablize the disk, preventing the process of star formation \citep[e.g.][]{Martig2009}. The more concentrated structures, higher halo masses and lower gas detection ratio together suggest NUV$-r$ selected MRSGs are in the later stage of MRSGs quenching. 

\begin{figure}
\includegraphics[scale=1]{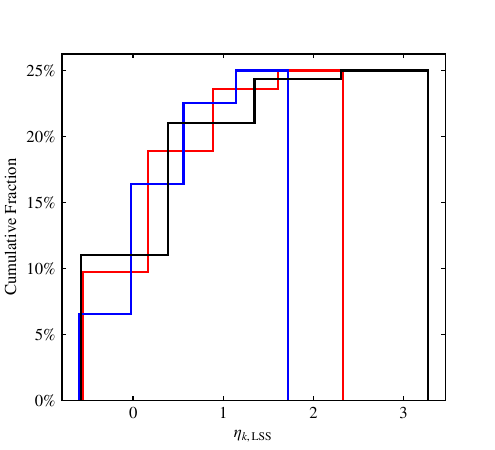}
\caption{The cumulative fraction of projected number density ($\eta_{k,\rm LSS}$) for different types of galaxies in our sample, where k=5 denotes the 5th nearest neighbor. MRSGs, blue spirals and S0s are shown in red, blue and black color. }
\label{eta}
\end{figure}

Figure \ref{eta} shows the cumulative distribution of $\eta_{k,\rm LSS}$ for different types of galaxies, with $\eta_{k,\rm LSS}$ representing the projected number density within the distance to the 5th nearest neighbor. This parameter is defines in \cite{Argudo2015}:

\begin{equation}
\eta_{k, \mathrm{LSS}} \equiv \log \left(\frac{k-1}{\operatorname{Vol}\left(d_k\right)}\right)=\log \left(\frac{3(k-1)}{4 \pi d_k^3}\right)
\end{equation}

We find S0s reside in the densest environment, while red spirals occupy a denser environment compared to blue spirals. We calculate values of the K–S statistics probability of $\eta_{5,\rm LSS}$ for MRSGs compared to S0s and blue spirals, which are 0.75 and 0.05 respectively. The environment of red spirals is different from blue spirals at a significance level >95\%.

\subsection{The Evolutionary Connection between MRSGs and S0s}

Through the analysis in Sec \ref{sec:spectroscopic} and Sec \ref{sec:structure}, we find MRSGs and S0 galaxies show similar distributions in multiple parameter spaces. This indicates the evolutionary connection exits between these two types of galaxies. Shock heating in massive halos prevent MRSGs from accreting fresh gas from surroundings. As the residual star formation in MRSGs is exhausted and the spiral arms fade away, MRSGs will evolve into S0s.

Simulations always suggest spiral arms are transient and recurrent, which is regulated by the balance between shear and self-gravity \citep{Carlberg1985,Bottema2003,Sellwood2011,Sellwood2013,Fujii2011ApJ,Grand2012,Grand2012b,Baba2013ApJ,Onghia2013ApJ,Roca2013}. 
However, modern simulations have revealed the non-linear effects in the formation of spiral arms \citep[e.g.][]{Onghia2013ApJ}. The waves triggered by the perturbers can travel fast and form new perturbers that continuously excite the disk. These non-linear effects allows for the maintainance of spiral structures for extended periods, even in the absence of new triggers. 

As mentioned in \citep{Onghia2013ApJ}, the non-linear effects can explain the presence of spiral structures in red spirals, which lack molecular clouds. 
Additionally, their results reveal that the lifetime and masses of GMCs which serves as original perturbers affects the amplitude of spiral arms. 
The absence of spiral features in S0s may result from the properties of GMCs at the formation time of S0s. The perturbations may not be strong enough to generate long-lasting spiral features, or these features may have vanished.
Actually, observations have identified faded spiral features in S0 galaxies \citep[e.g.][]{DeGraaff2007}, and simulations illustrate a stage in the evolution of S0 galaxies where spiral features can emerge \citep[e.g.][]{Deeley2020}.

Previous simulation studies revealed that MRSGs have similar evolution trajectory as some of S0 galaxies through mergers \citep[e.g.][]{Deeley2021}. The gas in the central part of MRSGs has depleted and the star formation has been shifted to the outer regions of galaxies. Examples of S0 galaxies which undergo merger events also stay in the spiral phase for some time before the star formation has totally quenched.
Compaction can also produce galaxies with similar morphologies as S0s and MRSGs. Gas disks at high redshift undergo instability, resulting in the formation of blue nuggets. These blue nuggets then experience rapid quenching and transform into red nuggets. These galaxies are believed to be the possible progenitors of ellipticals \citep{Dekel2014MNRAS}. However, there are cases that dry mergers form envelop or disc structures around the condensed bulge \citep{Zolotov2015}, which may generate red spirals or S0s under different post-compaction conditions. In this case, the red spirals may evolve to S0s if the residual star formation quench.

Besides mergers, gas stripping via group inflall is another main pathway for the formation of S0 galaxies. Since the disk is not ruined, these S0s hold rotation supported motions. This process tends to happen in denser environment like galaxy clusters, which is not included in our work. 
MRSGs in massive halos may also experience halo stripping where halo gas is stripped during hydrodynamical interaction between the intracluster gas \citep{Bekki2002}. This process will furthur increase the Q parameter for galaxy disks and make the spiral arm structure less pronounced.
\cite{Shimakawa2022} studied the phase-space diagram for MRSGs to find their infalling trajectories towards the cluster center and found excess of passive spirals in the region where galaxies are expected to experience more than once core crossing. Their results support for the scenario that MRSGs in cluster environment can be formed by ram pressure stripping and as the less pronounced spiral arms vanish, these MRSGs may evolve to S0s.



\section{Conclusions}
\label{sec:conclusions}
In this work, we use the spatially resolved spectroscopic data from MaNGA DR17 observation to study the spectroscopic and structural properties of MRSGs.
We limit our samples to massive galaxies which are above $\rm 10^{10.5}M_{\odot}$, and select MRSGs according to their location on the $g-r$ color mass diagram. Blue spirals and S0s with same mass range and B/T ratio are selected as control samples. There are 131 MRSGs, 117 blue spirals and 112 S0s in our sample. 
The main conclusions are summarized as follows.

\begin{enumerate}
    \item We find the random motions are more prominent in MRSGs compared with blue spirals. MRSGs and S0 galaxies have similar Dn4000, metallicity and Mgb/<Fe> radial profiles, indicating they are composed of similar old stellar populations with rapid-formed bulges. The differences between the age and star formation timescale of MRSGs and blue spirals suggest MRSGs are not simply evolutionary remnants of nearby blue spirals. Based on the similarities between MRSGs and S0s, we discuss the possible evolutionary trend between MRSGs and S0s.
    \item We confirm the residual star formation which resides in the outer regions of MRSGs and refer to 0.8 $R_{\rm e}$ as the boundary outside which star-forming spaxels are concentrated. We find although the number of star-forming spaxels in MRSGs are not as high as that of blue spirals, it is still conspicuous compared with S0s. Besides, star-forming spaxels in MRSGs are below the spatially resolved star-forming main sequence, suggesting the relatively weak star formation activity. We stack the spectrum within 0.8 $R_{\rm e}$ and out of 0.8 $R_{\rm e}$ separately. The spectrum of MRSGs resembles that of S0s which consists of old stellar population. However, obvious H$\alpha$ emission can be seen from the outer regions of MRSGs. We estimate the corresponding specific star formation rate to be -11.33$\pm$0.08 yr$^{-1}$, which is roughly 1 dex lower than that of blue spirals. The stellar population fitting shows that MRSGs have a negative age gradient.
    \item We compare the structural properties for MRSGs and the control samples, and find MRSGs are more compact than blue spirals at each mass bin. The average sizes of MRSGs fall in between blue spirals and S0s. We investigate the distribution of $\Sigma_1$ (stellar mass surface density within the central 1 kpc), and find that most of the MRSGs have reached a characteristic density for early-type galaxies at their centers, suggesting possible dissipative process during bulge formation. The distributions of S$\acute{e}$rsic index and concentration index also confirm the fact that MRSGs have concentrated structures. MRSGs are less asymmetric, which indicates the spiral structures are vanishing gradually. 
    \item We check the spectrum profile of MRSGs with NUV$-r$>5 (which we call NUV$-r$ selected MRSGs) and there is no obvious H$\alpha$ emission in the outer region, which indicates they are towards totally quenched. The low \HI{} detection rate of NUV$-r$ selected MRSGs is consistent with this result. NUV$-r$ selected MRSGs hold more prominent bulges with B/T > 0.5. The statistics on other structural parameters also show these NUV$-r$ selected MRSGs have more concentrated structures. NUV$-r$ selected MRSGs tend to reside in more massive halos, they seem to be in a later stage of MRSGs star formation quenching.
\end{enumerate}

\section{Acknowledgement}
We thank the anonymous referee for critical comments and instructive suggestions, which improved the content and analysis significantly.

This work is supported by the National Natural Science Foundation of China (No. 12192222, 12192220 and 12121003).

Funding for the Sloan Digital Sky 
Survey IV has been provided by the 
Alfred P. Sloan Foundation, the U.S. 
Department of Energy Office of 
Science, and the Participating 
Institutions. 

SDSS-IV acknowledges support and resources from the Center for High Performance Computing  at the 
University of Utah. The SDSS 
website is www.sdss4.org.

SDSS-IV is managed by the 
Astrophysical Research Consortium 
for the Participating Institutions 
of the SDSS Collaboration including 
the Brazilian Participation Group, 
the Carnegie Institution for Science, 
Carnegie Mellon University, Center for 
Astrophysics | Harvard \& 
Smithsonian, the Chilean Participation 
Group, the French Participation Group, 
Instituto de Astrof\'isica de 
Canarias, The Johns Hopkins 
University, Kavli Institute for the 
Physics and Mathematics of the 
Universe (IPMU) / University of 
Tokyo, the Korean Participation Group, 
Lawrence Berkeley National Laboratory, 
Leibniz Institut f\"ur Astrophysik 
Potsdam (AIP),  Max-Planck-Institut 
f\"ur Astronomie (MPIA Heidelberg), 
Max-Planck-Institut f\"ur 
Astrophysik (MPA Garching), 
Max-Planck-Institut f\"ur 
Extraterrestrische Physik (MPE), 
National Astronomical Observatories of 
China, New Mexico State University, 
New York University, University of 
Notre Dame, Observat\'ario 
Nacional / MCTI, The Ohio State 
University, Pennsylvania State 
University, Shanghai 
Astronomical Observatory, United 
Kingdom Participation Group, 
Universidad Nacional Aut\'onoma 
de M\'exico, University of Arizona, 
University of Colorado Boulder, 
University of Oxford, University of 
Portsmouth, University of Utah, 
University of Virginia, University 
of Washington, University of 
Wisconsin, Vanderbilt University, 
and Yale University.

This publication makes use of data products from the Wide-field Infrared Survey Explorer, which is a joint project of the University of California, Los Angeles, and the Jet Propulsion Laboratory/California Institute of Technology, funded by the National Aeronautics and Space Administration.

The Legacy Surveys consist of three individual and complementary projects: the Dark Energy Camera Legacy Survey (DECaLS; Proposal ID \#2014B-0404; PIs: David Schlegel and Arjun Dey), the Beijing-Arizona Sky Survey (BASS; NOAO Prop. ID \#2015A-0801; PIs: Zhou Xu and Xiaohui Fan), and the Mayall z-band Legacy Survey (MzLS; Prop. ID \#2016A-0453; PI: Arjun Dey). DECaLS, BASS and MzLS together include data obtained, respectively, at the Blanco telescope, Cerro Tololo Inter-American Observatory, NSF’s NOIRLab; the Bok telescope, Steward Observatory, University of Arizona; and the Mayall telescope, Kitt Peak National Observatory, NOIRLab. Pipeline processing and analyses of the data were supported by NOIRLab and the Lawrence Berkeley National Laboratory (LBNL). The Legacy Surveys project is honored to be permitted to conduct astronomical research on Iolkam Du’ag (Kitt Peak), a mountain with particular significance to the Tohono O’odham Nation.

NOIRLab is operated by the Association of Universities for Research in Astronomy (AURA) under a cooperative agreement with the National Science Foundation. LBNL is managed by the Regents of the University of California under contract to the U.S. Department of Energy.

This project used data obtained with the Dark Energy Camera (DECam), which was constructed by the Dark Energy Survey (DES) collaboration. Funding for the DES Projects has been provided by the U.S. Department of Energy, the U.S. National Science Foundation, the Ministry of Science and Education of Spain, the Science and Technology Facilities Council of the United Kingdom, the Higher Education Funding Council for England, the National Center for Supercomputing Applications at the University of Illinois at Urbana-Champaign, the Kavli Institute of Cosmological Physics at the University of Chicago, Center for Cosmology and Astro-Particle Physics at the Ohio State University, the Mitchell Institute for Fundamental Physics and Astronomy at Texas A\&M University, Financiadora de Estudos e Projetos, Fundacao Carlos Chagas Filho de Amparo, Financiadora de Estudos e Projetos, Fundacao Carlos Chagas Filho de Amparo a Pesquisa do Estado do Rio de Janeiro, Conselho Nacional de Desenvolvimento Cientifico e Tecnologico and the Ministerio da Ciencia, Tecnologia e Inovacao, the Deutsche Forschungsgemeinschaft and the Collaborating Institutions in the Dark Energy Survey. The Collaborating Institutions are Argonne National Laboratory, the University of California at Santa Cruz, the University of Cambridge, Centro de Investigaciones Energeticas, Medioambientales y Tecnologicas-Madrid, the University of Chicago, University College London, the DES-Brazil Consortium, the University of Edinburgh, the Eidgenossische Technische Hochschule (ETH) Zurich, Fermi National Accelerator Laboratory, the University of Illinois at Urbana-Champaign, the Institut de Ciencies de l’Espai (IEEC/CSIC), the Institut de Fisica d’Altes Energies, Lawrence Berkeley National Laboratory, the Ludwig Maximilians Universitat Munchen and the associated Excellence Cluster Universe, the University of Michigan, NSF’s NOIRLab, the University of Nottingham, the Ohio State University, the University of Pennsylvania, the University of Portsmouth, SLAC National Accelerator Laboratory, Stanford University, the University of Sussex, and Texas A\&M University.

BASS is a key project of the Telescope Access Program (TAP), which has been funded by the National Astronomical Observatories of China, the Chinese Academy of Sciences (the Strategic Priority Research Program “The Emergence of Cosmological Structures” Grant \# XDB09000000), and the Special Fund for Astronomy from the Ministry of Finance. The BASS is also supported by the External Cooperation Program of Chinese Academy of Sciences (Grant \# 114A11KYSB20160057), and Chinese National Natural Science Foundation (Grant \# 12120101003, \# 11433005).

The Legacy Survey team makes use of data products from the Near-Earth Object Wide-field Infrared Survey Explorer (NEOWISE), which is a project of the Jet Propulsion Laboratory/California Institute of Technology. NEOWISE is funded by the National Aeronautics and Space Administration.

The Legacy Surveys imaging of the DESI footprint is supported by the Director, Office of Science, Office of High Energy Physics of the U.S. Department of Energy under Contract No. DE-AC02-05CH1123, by the National Energy Research Scientific Computing Center, a DOE Office of Science User Facility under the same contract; and by the U.S. National Science Foundation, Division of Astronomical Sciences under Contract No. AST-0950945 to NOAO.

\section{Data Availability}
The data underlying this article were accessed from MaNGA Data Release 17. The DAP and PIPE3D pipeline data are available at \href{https://www.sdss.org/dr17/manga/}{https://www.sdss.org/dr17/manga/}. \textsc{starlight} is available at \href{http://www.starlight.ufsc.br}{http://www.starlight.ufsc.br}. \textsc{indexf} is available at \href{http://research.iac.es/galeria/vazdekis/}{http://research.iac.es/galeria/vazdekis/}.
The DESI false-color images are available at \href{https://www.legacysurvey.org}{https://www.legacysurvey.org}.



\bibliographystyle{mnras}
\bibliography{example} 

\begin{thebibliography}{}
\makeatletter
\relax
\def\mn@urlcharsother{\let\do\@makeother \do\$\do\&\do\#\do\^\do\_\do\%\do\~}
\def\mn@doi{\begingroup\mn@urlcharsother \@ifnextchar [ {\mn@doi@} {\mn@doi@[]}}
\def\mn@doi@[#1]#2{\def\@tempa{#1}\ifx\@tempa\@empty \href {http://dx.doi.org/#2} {doi:#2}\else \href {http://dx.doi.org/#2} {#1}\fi \endgroup}
\def\mn@eprint#1#2{\mn@eprint@#1:#2::\@nil}
\def\mn@eprint@arXiv#1{\href {http://arxiv.org/abs/#1} {{\tt arXiv:#1}}}
\def\mn@eprint@dblp#1{\href {http://dblp.uni-trier.de/rec/bibtex/#1.xml} {dblp:#1}}
\def\mn@eprint@#1:#2:#3:#4\@nil{\def\@tempa {#1}\def\@tempb {#2}\def\@tempc {#3}\ifx \@tempc \@empty \let \@tempc \@tempb \let \@tempb \@tempa \fi \ifx \@tempb \@empty \def\@tempb {arXiv}\fi \@ifundefined {mn@eprint@\@tempb}{\@tempb:\@tempc}{\expandafter \expandafter \csname mn@eprint@\@tempb\endcsname \expandafter{\@tempc}}}

\bibitem[\protect\citeauthoryear{{Abdurro'uf} et~al.,}{{Abdurro'uf} et~al.}{2022}]{Abdurrouf2022ApJS}
{Abdurro'uf} et~al., 2022, \mn@doi [\apjs] {10.3847/1538-4365/ac4414}, \href {https://ui.adsabs.harvard.edu/abs/2022ApJS..259...35A} {259, 35}

\bibitem[\protect\citeauthoryear{{Argudo-Fern{\'a}ndez} et~al.,}{{Argudo-Fern{\'a}ndez} et~al.}{2015}]{Argudo2015}
{Argudo-Fern{\'a}ndez} M.,  et~al., 2015, \mn@doi [\aap] {10.1051/0004-6361/201526016}, \href {https://ui.adsabs.harvard.edu/abs/2015A&A...578A.110A} {578, A110}

\bibitem[\protect\citeauthoryear{{Athanassoula}, {Rodionov}, {Peschken}  \& {Lambert}}{{Athanassoula} et~al.}{2016}]{Athanassoula2016}
{Athanassoula} E.,  {Rodionov} S.~A.,  {Peschken} N.,   {Lambert} J.~C.,  2016, \mn@doi [\apj] {10.3847/0004-637X/821/2/90}, \href {https://ui.adsabs.harvard.edu/abs/2016ApJ...821...90A} {821, 90}

\bibitem[\protect\citeauthoryear{{Baba}, {Saitoh}  \& {Wada}}{{Baba} et~al.}{2013}]{Baba2013ApJ}
{Baba} J.,  {Saitoh} T.~R.,   {Wada} K.,  2013, \mn@doi [\apj] {10.1088/0004-637X/763/1/46}, \href {https://ui.adsabs.harvard.edu/abs/2013ApJ...763...46B} {763, 46}

\bibitem[\protect\citeauthoryear{{Baldry}, {Glazebrook}, {Brinkmann}, {Ivezi{\'c}}, {Lupton}, {Nichol}  \& {Szalay}}{{Baldry} et~al.}{2004}]{Baldry2004}
{Baldry} I.~K.,  {Glazebrook} K.,  {Brinkmann} J.,  {Ivezi{\'c}} {\v{Z}}.,  {Lupton} R.~H.,  {Nichol} R.~C.,   {Szalay} A.~S.,  2004, \mn@doi [\apj] {10.1086/380092}, \href {https://ui.adsabs.harvard.edu/abs/2004ApJ...600..681B} {600, 681}

\bibitem[\protect\citeauthoryear{{Baldry}, {Balogh}, {Bower}, {Glazebrook}, {Nichol}, {Bamford}  \& {Budavari}}{{Baldry} et~al.}{2006}]{Baldry2006}
{Baldry} I.~K.,  {Balogh} M.~L.,  {Bower} R.~G.,  {Glazebrook} K.,  {Nichol} R.~C.,  {Bamford} S.~P.,   {Budavari} T.,  2006, \mn@doi [\mnras] {10.1111/j.1365-2966.2006.11081.x}, \href {https://ui.adsabs.harvard.edu/abs/2006MNRAS.373..469B} {373, 469}

\bibitem[\protect\citeauthoryear{{Barro} et~al.,}{{Barro} et~al.}{2017}]{Barro2017}
{Barro} G.,  et~al., 2017, \mn@doi [\apj] {10.3847/1538-4357/aa6b05}, \href {https://ui.adsabs.harvard.edu/abs/2017ApJ...840...47B} {840, 47}

\bibitem[\protect\citeauthoryear{{Bekki}, {Couch}  \& {Shioya}}{{Bekki} et~al.}{2002}]{Bekki2002}
{Bekki} K.,  {Couch} W.~J.,   {Shioya} Y.,  2002, \mn@doi [\apj] {10.1086/342221}, \href {https://ui.adsabs.harvard.edu/abs/2002ApJ...577..651B} {577, 651}

\bibitem[\protect\citeauthoryear{{Belfiore} et~al.,}{{Belfiore} et~al.}{2019}]{Belfiore2019}
{Belfiore} F.,  et~al., 2019, \mn@doi [\aj] {10.3847/1538-3881/ab3e4e}, \href {https://ui.adsabs.harvard.edu/abs/2019AJ....158..160B} {158, 160}

\bibitem[\protect\citeauthoryear{{Bell} et~al.,}{{Bell} et~al.}{2004}]{Bell2004}
{Bell} E.~F.,  et~al., 2004, \mn@doi [\apj] {10.1086/420778}, \href {https://ui.adsabs.harvard.edu/abs/2004ApJ...608..752B} {608, 752}

\bibitem[\protect\citeauthoryear{{Bell} et~al.,}{{Bell} et~al.}{2012}]{Bell2012}
{Bell} E.~F.,  et~al., 2012, \mn@doi [\apj] {10.1088/0004-637X/753/2/167}, \href {https://ui.adsabs.harvard.edu/abs/2012ApJ...753..167B} {753, 167}

\bibitem[\protect\citeauthoryear{{Bertin} \& {Lin}}{{Bertin} \& {Lin}}{1996}]{Bertin1996}
{Bertin} G.,  {Lin} C.~C.,  1996, {Spiral structure in galaxies a density wave theory}

\bibitem[\protect\citeauthoryear{{Blanton} et~al.,}{{Blanton} et~al.}{2005}]{Blanton2005}
{Blanton} M.~R.,  et~al., 2005, \mn@doi [\aj] {10.1086/429803}, \href {https://ui.adsabs.harvard.edu/abs/2005AJ....129.2562B} {129, 2562}

\bibitem[\protect\citeauthoryear{{Blanton} et~al.,}{{Blanton} et~al.}{2017}]{Blanton2017}
{Blanton} M.~R.,  et~al., 2017, \mn@doi [\aj] {10.3847/1538-3881/aa7567}, \href {https://ui.adsabs.harvard.edu/abs/2017AJ....154...28B} {154, 28}

\bibitem[\protect\citeauthoryear{{Bluck}, {Mendel}, {Ellison}, {Moreno}, {Simard}, {Patton}  \& {Starkenburg}}{{Bluck} et~al.}{2014}]{Bluck2014}
{Bluck} A. F.~L.,  {Mendel} J.~T.,  {Ellison} S.~L.,  {Moreno} J.,  {Simard} L.,  {Patton} D.~R.,   {Starkenburg} E.,  2014, \mn@doi [\mnras] {10.1093/mnras/stu594}, \href {https://ui.adsabs.harvard.edu/abs/2014MNRAS.441..599B} {441, 599}

\bibitem[\protect\citeauthoryear{{Bottema}}{{Bottema}}{2003}]{Bottema2003}
{Bottema} R.,  2003, \mn@doi [\mnras] {10.1046/j.1365-8711.2003.06613.x}, \href {https://ui.adsabs.harvard.edu/abs/2003MNRAS.344..358B} {344, 358}

\bibitem[\protect\citeauthoryear{{Brinchmann}, {Charlot}, {White}, {Tremonti}, {Kauffmann}, {Heckman}  \& {Brinkmann}}{{Brinchmann} et~al.}{2004}]{Brinchmann2004}
{Brinchmann} J.,  {Charlot} S.,  {White} S.~D.~M.,  {Tremonti} C.,  {Kauffmann} G.,  {Heckman} T.,   {Brinkmann} J.,  2004, \mn@doi [\mnras] {10.1111/j.1365-2966.2004.07881.x}, \href {https://ui.adsabs.harvard.edu/abs/2004MNRAS.351.1151B} {351, 1151}

\bibitem[\protect\citeauthoryear{{Bruzual} \& {Charlot}}{{Bruzual} \& {Charlot}}{2003}]{Bruzual2003}
{Bruzual} G.,  {Charlot} S.,  2003, \mn@doi [\mnras] {10.1046/j.1365-8711.2003.06897.x}, \href {https://ui.adsabs.harvard.edu/abs/2003MNRAS.344.1000B} {344, 1000}

\bibitem[\protect\citeauthoryear{{Bundy} et~al.,}{{Bundy} et~al.}{2010}]{Bundy2010}
{Bundy} K.,  et~al., 2010, \mn@doi [\apj] {10.1088/0004-637X/719/2/1969}, \href {https://ui.adsabs.harvard.edu/abs/2010ApJ...719.1969B} {719, 1969}

\bibitem[\protect\citeauthoryear{{Bundy} et~al.,}{{Bundy} et~al.}{2015}]{Bundy2015}
{Bundy} K.,  et~al., 2015, \mn@doi [\apj] {10.1088/0004-637X/798/1/7}, \href {https://ui.adsabs.harvard.edu/abs/2015ApJ...798....7B} {798, 7}

\bibitem[\protect\citeauthoryear{{Cappellari} et~al.,}{{Cappellari} et~al.}{2007}]{Cappellari2007}
{Cappellari} M.,  et~al., 2007, \mn@doi [\mnras] {10.1111/j.1365-2966.2007.11963.x}, \href {https://ui.adsabs.harvard.edu/abs/2007MNRAS.379..418C} {379, 418}

\bibitem[\protect\citeauthoryear{{Cardiel}}{{Cardiel}}{2010}]{Cardiel2010}
{Cardiel} N.,  2010, {indexf: Line-strength Indices in Fully Calibrated FITS Spectra}, Astrophysics Source Code Library, record ascl:1010.046 (\mn@eprint {ascl} {1010.046})

\bibitem[\protect\citeauthoryear{{Cardiel}, {Gorgas}, {Cenarro}  \& {Gonzalez}}{{Cardiel} et~al.}{1998}]{Cardiel1998}
{Cardiel} N.,  {Gorgas} J.,  {Cenarro} J.,   {Gonzalez} J.~J.,  1998, \mn@doi [\aaps] {10.1051/aas:1998123}, \href {https://ui.adsabs.harvard.edu/abs/1998A&AS..127..597C} {127, 597}

\bibitem[\protect\citeauthoryear{{Carlberg} \& {Freedman}}{{Carlberg} \& {Freedman}}{1985}]{Carlberg1985}
{Carlberg} R.~G.,  {Freedman} W.~L.,  1985, \mn@doi [\apj] {10.1086/163634}, \href {https://ui.adsabs.harvard.edu/abs/1985ApJ...298..486C} {298, 486}

\bibitem[\protect\citeauthoryear{{Carson} \& {Nichol}}{{Carson} \& {Nichol}}{2010}]{Carson2010}
{Carson} D.~P.,  {Nichol} R.~C.,  2010, \mn@doi [\mnras] {10.1111/j.1365-2966.2010.17151.x}, \href {https://ui.adsabs.harvard.edu/abs/2010MNRAS.408..213C} {408, 213}

\bibitem[\protect\citeauthoryear{{Chabrier}}{{Chabrier}}{2003}]{Chabrier2003}
{Chabrier} G.,  2003, \mn@doi [\pasp] {10.1086/376392}, \href {https://ui.adsabs.harvard.edu/abs/2003PASP..115..763C} {115, 763}

\bibitem[\protect\citeauthoryear{{Chen}, {Lowenthal}  \& {Yun}}{{Chen} et~al.}{2010}]{Chen2010}
{Chen} Y.,  {Lowenthal} J.~D.,   {Yun} M.~S.,  2010, \mn@doi [\apj] {10.1088/0004-637X/712/2/1385}, \href {https://ui.adsabs.harvard.edu/abs/2010ApJ...712.1385C} {712, 1385}

\bibitem[\protect\citeauthoryear{{Chen} et~al.,}{{Chen} et~al.}{2020}]{Chen2020ApJ}
{Chen} Z.,  et~al., 2020, \mn@doi [\apj] {10.3847/1538-4357/ab9633}, \href {https://ui.adsabs.harvard.edu/abs/2020ApJ...897..102C} {897, 102}

\bibitem[\protect\citeauthoryear{{Cheung} et~al.,}{{Cheung} et~al.}{2012}]{Cheung2012}
{Cheung} E.,  et~al., 2012, \mn@doi [\apj] {10.1088/0004-637X/760/2/131}, \href {https://ui.adsabs.harvard.edu/abs/2012ApJ...760..131C} {760, 131}

\bibitem[\protect\citeauthoryear{{Cid Fernandes}, {Mateus}, {Sodr{\'e}}, {Stasi{\'n}ska}  \& {Gomes}}{{Cid Fernandes} et~al.}{2005}]{CidFernandes2005}
{Cid Fernandes} R.,  {Mateus} A.,  {Sodr{\'e}} L.,  {Stasi{\'n}ska} G.,   {Gomes} J.~M.,  2005, \mn@doi [\mnras] {10.1111/j.1365-2966.2005.08752.x}, \href {https://ui.adsabs.harvard.edu/abs/2005MNRAS.358..363C} {358, 363}

\bibitem[\protect\citeauthoryear{{Cid Fernandes}, {Stasi{\'n}ska}, {Mateus}  \& {Vale Asari}}{{Cid Fernandes} et~al.}{2011}]{Fernandes2011}
{Cid Fernandes} R.,  {Stasi{\'n}ska} G.,  {Mateus} A.,   {Vale Asari} N.,  2011, \mn@doi [\mnras] {10.1111/j.1365-2966.2011.18244.x}, \href {https://ui.adsabs.harvard.edu/abs/2011MNRAS.413.1687C} {413, 1687}

\bibitem[\protect\citeauthoryear{{Combes} \& {Sanders}}{{Combes} \& {Sanders}}{1981}]{CS1981}
{Combes} F.,  {Sanders} R.~H.,  1981, \aap, \href {https://ui.adsabs.harvard.edu/abs/1981A&A....96..164C} {96, 164}

\bibitem[\protect\citeauthoryear{{Conselice}}{{Conselice}}{2006}]{Conselice2006}
{Conselice} C.~J.,  2006, \mn@doi [\mnras] {10.1111/j.1365-2966.2006.11114.x}, \href {https://ui.adsabs.harvard.edu/abs/2006MNRAS.373.1389C} {373, 1389}

\bibitem[\protect\citeauthoryear{{Cortese}}{{Cortese}}{2012}]{Cortese2012}
{Cortese} L.,  2012, \mn@doi [\aap] {10.1051/0004-6361/201219443}, \href {https://ui.adsabs.harvard.edu/abs/2012A&A...543A.132C} {543, A132}

\bibitem[\protect\citeauthoryear{{Crowl}, {Kenney}, {van Gorkom}  \& {Vollmer}}{{Crowl} et~al.}{2005}]{Crowl2005}
{Crowl} H.~H.,  {Kenney} J. D.~P.,  {van Gorkom} J.~H.,   {Vollmer} B.,  2005, \mn@doi [\aj] {10.1086/430526}, \href {https://ui.adsabs.harvard.edu/abs/2005AJ....130...65C} {130, 65}

\bibitem[\protect\citeauthoryear{{D'Onghia}, {Vogelsberger}  \& {Hernquist}}{{D'Onghia} et~al.}{2013}]{Onghia2013ApJ}
{D'Onghia} E.,  {Vogelsberger} M.,   {Hernquist} L.,  2013, \mn@doi [\apj] {10.1088/0004-637X/766/1/34}, \href {https://ui.adsabs.harvard.edu/abs/2013ApJ...766...34D} {766, 34}

\bibitem[\protect\citeauthoryear{{DeGraaff}, {Blakeslee}, {Meurer}  \& {Putman}}{{DeGraaff} et~al.}{2007}]{DeGraaff2007}
{DeGraaff} R.~B.,  {Blakeslee} J.~P.,  {Meurer} G.~R.,   {Putman} M.~E.,  2007, \mn@doi [\apj] {10.1086/523640}, \href {https://ui.adsabs.harvard.edu/abs/2007ApJ...671.1624D} {671, 1624}

\bibitem[\protect\citeauthoryear{{Deeley} et~al.,}{{Deeley} et~al.}{2020}]{Deeley2020}
{Deeley} S.,  et~al., 2020, \mn@doi [\mnras] {10.1093/mnras/staa2417}, \href {https://ui.adsabs.harvard.edu/abs/2020MNRAS.498.2372D} {498, 2372}

\bibitem[\protect\citeauthoryear{{Deeley}, {Drinkwater}, {Sweet}, {Bekki}, {Couch}, {Forbes}  \& {Dolfi}}{{Deeley} et~al.}{2021}]{Deeley2021}
{Deeley} S.,  {Drinkwater} M.~J.,  {Sweet} S.~M.,  {Bekki} K.,  {Couch} W.~J.,  {Forbes} D.~A.,   {Dolfi} A.,  2021, \mn@doi [\mnras] {10.1093/mnras/stab2007}, \href {https://ui.adsabs.harvard.edu/abs/2021MNRAS.508..895D} {508, 895}

\bibitem[\protect\citeauthoryear{{Dekel} \& {Birnboim}}{{Dekel} \& {Birnboim}}{2006}]{Dekel2006}
{Dekel} A.,  {Birnboim} Y.,  2006, \mn@doi [\mnras] {10.1111/j.1365-2966.2006.10145.x}, \href {https://ui.adsabs.harvard.edu/abs/2006MNRAS.368....2D} {368, 2}

\bibitem[\protect\citeauthoryear{{Dekel} \& {Burkert}}{{Dekel} \& {Burkert}}{2014}]{Dekel2014MNRAS}
{Dekel} A.,  {Burkert} A.,  2014, \mn@doi [\mnras] {10.1093/mnras/stt2331}, \href {https://ui.adsabs.harvard.edu/abs/2014MNRAS.438.1870D} {438, 1870}

\bibitem[\protect\citeauthoryear{{Dey} et~al.,}{{Dey} et~al.}{2019}]{Dey2019AJ}
{Dey} A.,  et~al., 2019, \mn@doi [\aj] {10.3847/1538-3881/ab089d}, \href {https://ui.adsabs.harvard.edu/abs/2019AJ....157..168D} {157, 168}

\bibitem[\protect\citeauthoryear{{\VAN{Dijk}{Van}{van}}~Dijk}{{\VAN{Dijk}{Van}{van}}~Dijk}{1902}]{vanDijk1902}
{\VAN{Dijk}{Van}{van}}~Dijk T.,  1902, QJRAS, 2, 202

\bibitem[\protect\citeauthoryear{{Dom{\'\i}nguez S{\'a}nchez}, {Margalef}, {Bernardi}  \& {Huertas-Company}}{{Dom{\'\i}nguez S{\'a}nchez} et~al.}{2022}]{Dominguez2022}
{Dom{\'\i}nguez S{\'a}nchez} H.,  {Margalef} B.,  {Bernardi} M.,   {Huertas-Company} M.,  2022, \mn@doi [\mnras] {10.1093/mnras/stab3089}, \href {https://ui.adsabs.harvard.edu/abs/2022MNRAS.509.4024D} {509, 4024}

\bibitem[\protect\citeauthoryear{{Dressler}, {Smail}, {Poggianti}, {Butcher}, {Couch}, {Ellis}  \& {Oemler}}{{Dressler} et~al.}{1999}]{Dressler1999}
{Dressler} A.,  {Smail} I.,  {Poggianti} B.~M.,  {Butcher} H.,  {Couch} W.~J.,  {Ellis} R.~S.,   {Oemler} Augustus J.,  1999, \mn@doi [\apjs] {10.1086/313213}, \href {https://ui.adsabs.harvard.edu/abs/1999ApJS..122...51D} {122, 51}

\bibitem[\protect\citeauthoryear{{Dutton}, {van den Bosch}, {Dekel}  \& {Courteau}}{{Dutton} et~al.}{2007}]{Dutton2007}
{Dutton} A.~A.,  {van den Bosch} F.~C.,  {Dekel} A.,   {Courteau} S.,  2007, \mn@doi [\apj] {10.1086/509314}, \href {https://ui.adsabs.harvard.edu/abs/2007ApJ...654...27D} {654, 27}

\bibitem[\protect\citeauthoryear{{Enia} et~al.,}{{Enia} et~al.}{2020}]{Enia2020}
{Enia} A.,  et~al., 2020, \mn@doi [\mnras] {10.1093/mnras/staa433}, \href {https://ui.adsabs.harvard.edu/abs/2020MNRAS.493.4107E} {493, 4107}

\bibitem[\protect\citeauthoryear{{Etherington} \& {Thomas}}{{Etherington} \& {Thomas}}{2015}]{Etherington2015}
{Etherington} J.,  {Thomas} D.,  2015, \mn@doi [\mnras] {10.1093/mnras/stv999}, \href {https://ui.adsabs.harvard.edu/abs/2015MNRAS.451..660E} {451, 660}

\bibitem[\protect\citeauthoryear{{Faber} et~al.,}{{Faber} et~al.}{2007}]{Fabor2007}
{Faber} S.~M.,  et~al., 2007, \mn@doi [\apj] {10.1086/519294}, \href {https://ui.adsabs.harvard.edu/abs/2007ApJ...665..265F} {665, 265}

\bibitem[\protect\citeauthoryear{{Fall} \& {Efstathiou}}{{Fall} \& {Efstathiou}}{1980}]{Fall&Efstathiou1980}
{Fall} S.~M.,  {Efstathiou} G.,  1980, \mn@doi [\mnras] {10.1093/mnras/193.2.189}, \href {https://ui.adsabs.harvard.edu/abs/1980MNRAS.193..189F} {193, 189}

\bibitem[\protect\citeauthoryear{{Fang}, {Faber}, {Koo}  \& {Dekel}}{{Fang} et~al.}{2013}]{Fang2013}
{Fang} J.~J.,  {Faber} S.~M.,  {Koo} D.~C.,   {Dekel} A.,  2013, \mn@doi [\apj] {10.1088/0004-637X/776/1/63}, \href {https://ui.adsabs.harvard.edu/abs/2013ApJ...776...63F} {776, 63}

\bibitem[\protect\citeauthoryear{{Fogarty} et~al.,}{{Fogarty} et~al.}{2015}]{Fogarty2015}
{Fogarty} L.~M.~R.,  et~al., 2015, \mn@doi [\mnras] {10.1093/mnras/stv2060}, \href {https://ui.adsabs.harvard.edu/abs/2015MNRAS.454.2050F} {454, 2050}

\bibitem[\protect\citeauthoryear{Fournier}{Fournier}{1901}]{Fournier1901}
Fournier P.,  1901, ApJ, 1, 101

\bibitem[\protect\citeauthoryear{{Franx}, {van Dokkum}, {F{\"o}rster Schreiber}, {Wuyts}, {Labb{\'e}}  \& {Toft}}{{Franx} et~al.}{2008}]{Franx2008}
{Franx} M.,  {van Dokkum} P.~G.,  {F{\"o}rster Schreiber} N.~M.,  {Wuyts} S.,  {Labb{\'e}} I.,   {Toft} S.,  2008, \mn@doi [\apj] {10.1086/592431}, \href {https://ui.adsabs.harvard.edu/abs/2008ApJ...688..770F} {688, 770}

\bibitem[\protect\citeauthoryear{{Fraser-McKelvie}, {Brown}, {Pimbblet}, {Dolley}  \& {Bonne}}{{Fraser-McKelvie} et~al.}{2018}]{Fraser-McKelvie2018}
{Fraser-McKelvie} A.,  {Brown} M. J.~I.,  {Pimbblet} K.,  {Dolley} T.,   {Bonne} N.~J.,  2018, \mn@doi [\mnras] {10.1093/mnras/stx2823}, \href {https://ui.adsabs.harvard.edu/abs/2018MNRAS.474.1909F} {474, 1909}

\bibitem[\protect\citeauthoryear{{Fujii}, {Baba}, {Saitoh}, {Makino}, {Kokubo}  \& {Wada}}{{Fujii} et~al.}{2011}]{Fujii2011ApJ}
{Fujii} M.~S.,  {Baba} J.,  {Saitoh} T.~R.,  {Makino} J.,  {Kokubo} E.,   {Wada} K.,  2011, \mn@doi [\apj] {10.1088/0004-637X/730/2/109}, \href {https://ui.adsabs.harvard.edu/abs/2011ApJ...730..109F} {730, 109}

\bibitem[\protect\citeauthoryear{{Giovanelli} et~al.,}{{Giovanelli} et~al.}{2005}]{Giovanelli2005}
{Giovanelli} R.,  et~al., 2005, \mn@doi [\aj] {10.1086/497431}, \href {https://ui.adsabs.harvard.edu/abs/2005AJ....130.2598G} {130, 2598}

\bibitem[\protect\citeauthoryear{{Goldreich} \& {Lynden-Bell}}{{Goldreich} \& {Lynden-Bell}}{1965}]{Goldreich1965}
{Goldreich} P.,  {Lynden-Bell} D.,  1965, \mn@doi [\mnras] {10.1093/mnras/130.2.125}, \href {https://ui.adsabs.harvard.edu/abs/1965MNRAS.130..125G} {130, 125}

\bibitem[\protect\citeauthoryear{{Goto}, {Yamauchi}, {Fujita}, {Okamura}, {Sekiguchi}, {Smail}, {Bernardi}  \& {Gomez}}{{Goto} et~al.}{2003}]{Goto2003}
{Goto} T.,  {Yamauchi} C.,  {Fujita} Y.,  {Okamura} S.,  {Sekiguchi} M.,  {Smail} I.,  {Bernardi} M.,   {Gomez} P.~L.,  2003, \mn@doi [\mnras] {10.1046/j.1365-2966.2003.07114.x}, \href {https://ui.adsabs.harvard.edu/abs/2003MNRAS.346..601G} {346, 601}

\bibitem[\protect\citeauthoryear{{Grand}, {Kawata}  \& {Cropper}}{{Grand} et~al.}{2012a}]{Grand2012b}
{Grand} R. J.~J.,  {Kawata} D.,   {Cropper} M.,  2012a, \mn@doi [\mnras] {10.1111/j.1365-2966.2012.20411.x}, \href {https://ui.adsabs.harvard.edu/abs/2012MNRAS.421.1529G} {421, 1529}

\bibitem[\protect\citeauthoryear{{Grand}, {Kawata}  \& {Cropper}}{{Grand} et~al.}{2012b}]{Grand2012}
{Grand} R. J.~J.,  {Kawata} D.,   {Cropper} M.,  2012b, \mn@doi [\mnras] {10.1111/j.1365-2966.2012.21733.x}, \href {https://ui.adsabs.harvard.edu/abs/2012MNRAS.426..167G} {426, 167}

\bibitem[\protect\citeauthoryear{{\VAN{Guarde}{De la}{de la}}~Guarde}{{\VAN{Guarde}{De la}{de la}}~Guarde}{1904}]{delaGuarde1904}
{\VAN{Guarde}{De la}{de la}}~Guarde S.,  1904, MNRAS, 4, 404

\bibitem[\protect\citeauthoryear{{Guo}, {Hao}, {Xia}, {Shi}, {Chen}, {Li}  \& {Gu}}{{Guo} et~al.}{2020}]{Guo2020}
{Guo} R.,  {Hao} C.-N.,  {Xia} X.,  {Shi} Y.,  {Chen} Y.,  {Li} S.,   {Gu} Q.,  2020, \mn@doi [\apj] {10.3847/1538-4357/ab9b75}, \href {https://ui.adsabs.harvard.edu/abs/2020ApJ...897..162G} {897, 162}

\bibitem[\protect\citeauthoryear{{Hao}, {Shi}, {Chen}, {Xia}, {Gu}, {Guo}, {Yu}  \& {Li}}{{Hao} et~al.}{2019}]{Hao2019}
{Hao} C.-N.,  {Shi} Y.,  {Chen} Y.,  {Xia} X.,  {Gu} Q.,  {Guo} R.,  {Yu} X.,   {Li} S.,  2019, \mn@doi [\apjl] {10.3847/2041-8213/ab42e5}, \href {https://ui.adsabs.harvard.edu/abs/2019ApJ...883L..36H} {883, L36}

\bibitem[\protect\citeauthoryear{{Haynes} et~al.,}{{Haynes} et~al.}{2018}]{Haynes2018}
{Haynes} M.~P.,  et~al., 2018, \mn@doi [\apj] {10.3847/1538-4357/aac956}, \href {https://ui.adsabs.harvard.edu/abs/2018ApJ...861...49H} {861, 49}

\bibitem[\protect\citeauthoryear{{Hopkins}, {Cox}, {Younger}  \& {Hernquist}}{{Hopkins} et~al.}{2009a}]{Hopkins2009}
{Hopkins} P.~F.,  {Cox} T.~J.,  {Younger} J.~D.,   {Hernquist} L.,  2009a, \mn@doi [\apj] {10.1088/0004-637X/691/2/1168}, \href {https://ui.adsabs.harvard.edu/abs/2009ApJ...691.1168H} {691, 1168}

\bibitem[\protect\citeauthoryear{{Hopkins}, {Hernquist}, {Cox}, {Keres}  \& {Wuyts}}{{Hopkins} et~al.}{2009b}]{Hopkins2009ApJ}
{Hopkins} P.~F.,  {Hernquist} L.,  {Cox} T.~J.,  {Keres} D.,   {Wuyts} S.,  2009b, \mn@doi [\apj] {10.1088/0004-637X/691/2/1424}, \href {https://ui.adsabs.harvard.edu/abs/2009ApJ...691.1424H} {691, 1424}

\bibitem[\protect\citeauthoryear{{Hubble}}{{Hubble}}{1936}]{Hubble1936}
{Hubble} E.~P.,  1936, {Realm of the Nebulae}

\bibitem[\protect\citeauthoryear{{Ilbert} et~al.,}{{Ilbert} et~al.}{2010}]{Ilbert2010}
{Ilbert} O.,  et~al., 2010, \mn@doi [\apj] {10.1088/0004-637X/709/2/644}, \href {https://ui.adsabs.harvard.edu/abs/2010ApJ...709..644I} {709, 644}

\bibitem[\protect\citeauthoryear{{Jarrett} et~al.,}{{Jarrett} et~al.}{2011}]{Jarrett2011ApJ}
{Jarrett} T.~H.,  et~al., 2011, \mn@doi [\apj] {10.1088/0004-637X/735/2/112}, \href {https://ui.adsabs.harvard.edu/abs/2011ApJ...735..112J} {735, 112}

\bibitem[\protect\citeauthoryear{{Julian} \& {Toomre}}{{Julian} \& {Toomre}}{1966}]{Julian1966}
{Julian} W.~H.,  {Toomre} A.,  1966, \mn@doi [\apj] {10.1086/148957}, \href {https://ui.adsabs.harvard.edu/abs/1966ApJ...146..810J} {146, 810}

\bibitem[\protect\citeauthoryear{{Kauffmann} et~al.,}{{Kauffmann} et~al.}{2003a}]{Kauffmann2003b}
{Kauffmann} G.,  et~al., 2003a, \mn@doi [\mnras] {10.1046/j.1365-8711.2003.06291.x}, \href {https://ui.adsabs.harvard.edu/abs/2003MNRAS.341...33K} {341, 33}

\bibitem[\protect\citeauthoryear{{Kauffmann} et~al.,}{{Kauffmann} et~al.}{2003b}]{Kauffmann2003}
{Kauffmann} G.,  et~al., 2003b, \mn@doi [\mnras] {10.1046/j.1365-8711.2003.06291.x}, \href {https://ui.adsabs.harvard.edu/abs/2003MNRAS.341...33K} {341, 33}

\bibitem[\protect\citeauthoryear{{Kauffmann}, {Heckman}, {De Lucia}, {Brinchmann}, {Charlot}, {Tremonti}, {White}  \& {Brinkmann}}{{Kauffmann} et~al.}{2006}]{Kauffmann2006}
{Kauffmann} G.,  {Heckman} T.~M.,  {De Lucia} G.,  {Brinchmann} J.,  {Charlot} S.,  {Tremonti} C.,  {White} S. D.~M.,   {Brinkmann} J.,  2006, \mn@doi [\mnras] {10.1111/j.1365-2966.2006.10061.x}, \href {https://ui.adsabs.harvard.edu/abs/2006MNRAS.367.1394K} {367, 1394}

\bibitem[\protect\citeauthoryear{{Kennicutt}}{{Kennicutt}}{1998}]{Kennicutt1998}
{Kennicutt} Robert~C. J.,  1998, \mn@doi [\araa] {10.1146/annurev.astro.36.1.189}, \href {https://ui.adsabs.harvard.edu/abs/1998ARA&A..36..189K} {36, 189}

\bibitem[\protect\citeauthoryear{{Kobayashi}}{{Kobayashi}}{2004}]{Kobayashi2004}
{Kobayashi} C.,  2004, \mn@doi [\mnras] {10.1111/j.1365-2966.2004.07258.x}, \href {https://ui.adsabs.harvard.edu/abs/2004MNRAS.347..740K} {347, 740}

\bibitem[\protect\citeauthoryear{{Krajnovi{\'c}} et~al.,}{{Krajnovi{\'c}} et~al.}{2013}]{Krajnovic2013}
{Krajnovi{\'c}} D.,  et~al., 2013, \mn@doi [\mnras] {10.1093/mnras/sts315}, \href {https://ui.adsabs.harvard.edu/abs/2013MNRAS.432.1768K} {432, 1768}

\bibitem[\protect\citeauthoryear{{Kroupa}}{{Kroupa}}{2001}]{Kroupa2001}
{Kroupa} P.,  2001, \mn@doi [\mnras] {10.1046/j.1365-8711.2001.04022.x}, \href {https://ui.adsabs.harvard.edu/abs/2001MNRAS.322..231K} {322, 231}

\bibitem[\protect\citeauthoryear{{\VAN{Laguarde}{De}{de}}~Laguarde}{{\VAN{Laguarde}{De}{de}}~Laguarde}{1903}]{deLaguarde1903}
{\VAN{Laguarde}{De}{de}}~Laguarde A.,  1903, Nat, 3, 303

\bibitem[\protect\citeauthoryear{{Law} et~al.,}{{Law} et~al.}{2016}]{Law2016}
{Law} D.~R.,  et~al., 2016, \mn@doi [\aj] {10.3847/0004-6256/152/4/83}, \href {https://ui.adsabs.harvard.edu/abs/2016AJ....152...83L} {152, 83}

\bibitem[\protect\citeauthoryear{{Lin} \& {Shu}}{{Lin} \& {Shu}}{1964}]{Lin1964}
{Lin} C.~C.,  {Shu} F.~H.,  1964, \mn@doi [\apj] {10.1086/147955}, \href {https://ui.adsabs.harvard.edu/abs/1964ApJ...140..646L} {140, 646}

\bibitem[\protect\citeauthoryear{{Mahajan} et~al.,}{{Mahajan} et~al.}{2020}]{Mahajan2020}
{Mahajan} S.,  et~al., 2020, \mn@doi [\mnras] {10.1093/mnras/stz2993}, \href {https://ui.adsabs.harvard.edu/abs/2020MNRAS.491..398M} {491, 398}

\bibitem[\protect\citeauthoryear{{Marchesini} et~al.,}{{Marchesini} et~al.}{2014}]{Marchesini2014}
{Marchesini} D.,  et~al., 2014, \mn@doi [\apj] {10.1088/0004-637X/794/1/65}, \href {https://ui.adsabs.harvard.edu/abs/2014ApJ...794...65M} {794, 65}

\bibitem[\protect\citeauthoryear{{Martig}, {Bournaud}, {Teyssier}  \& {Dekel}}{{Martig} et~al.}{2009}]{Martig2009}
{Martig} M.,  {Bournaud} F.,  {Teyssier} R.,   {Dekel} A.,  2009, \mn@doi [\apj] {10.1088/0004-637X/707/1/250}, \href {https://ui.adsabs.harvard.edu/abs/2009ApJ...707..250M} {707, 250}

\bibitem[\protect\citeauthoryear{{Martin} et~al.,}{{Martin} et~al.}{2005}]{Martin2005}
{Martin} D.~C.,  et~al., 2005, \mn@doi [\apjl] {10.1086/426387}, \href {https://ui.adsabs.harvard.edu/abs/2005ApJ...619L...1M} {619, L1}

\bibitem[\protect\citeauthoryear{{Masters} et~al.,}{{Masters} et~al.}{2010}]{Masters2010}
{Masters} K.~L.,  et~al., 2010, \mn@doi [\mnras] {10.1111/j.1365-2966.2010.16503.x}, \href {https://ui.adsabs.harvard.edu/abs/2010MNRAS.405..783M} {405, 783}

\bibitem[\protect\citeauthoryear{{Masters} et~al.,}{{Masters} et~al.}{2019}]{Masters2019}
{Masters} K.~L.,  et~al., 2019, \mn@doi [\mnras] {10.1093/mnras/stz1889}, \href {https://ui.adsabs.harvard.edu/abs/2019MNRAS.488.3396M} {488, 3396}

\bibitem[\protect\citeauthoryear{{Matteucci} \& {Greggio}}{{Matteucci} \& {Greggio}}{1986}]{Matteucci1986}
{Matteucci} F.,  {Greggio} L.,  1986, \aap, \href {https://ui.adsabs.harvard.edu/abs/1986A&A...154..279M} {154, 279}

\bibitem[\protect\citeauthoryear{{Meert}, {Vikram}  \& {Bernardi}}{{Meert} et~al.}{2015}]{Meert2015}
{Meert} A.,  {Vikram} V.,   {Bernardi} M.,  2015, \mn@doi [\mnras] {10.1093/mnras/stu2333}, \href {https://ui.adsabs.harvard.edu/abs/2015MNRAS.446.3943M} {446, 3943}

\bibitem[\protect\citeauthoryear{{Mendel}, {Simard}, {Palmer}, {Ellison}  \& {Patton}}{{Mendel} et~al.}{2014}]{Mendel2014}
{Mendel} J.~T.,  {Simard} L.,  {Palmer} M.,  {Ellison} S.~L.,   {Patton} D.~R.,  2014, \mn@doi [\apjs] {10.1088/0067-0049/210/1/3}, \href {https://ui.adsabs.harvard.edu/abs/2014ApJS..210....3M} {210, 3}

\bibitem[\protect\citeauthoryear{{Mo}, {Mao}  \& {White}}{{Mo} et~al.}{1998}]{Mo1998}
{Mo} H.~J.,  {Mao} S.,   {White} S. D.~M.,  1998, \mn@doi [\mnras] {10.1046/j.1365-8711.1998.01227.x}, \href {https://ui.adsabs.harvard.edu/abs/1998MNRAS.295..319M} {295, 319}

\bibitem[\protect\citeauthoryear{{Mu{\~n}oz-Mateos}, {Gil de Paz}, {Boissier}, {Zamorano}, {Jarrett}, {Gallego}  \& {Madore}}{{Mu{\~n}oz-Mateos} et~al.}{2007a}]{Munoz-Mateos2007}
{Mu{\~n}oz-Mateos} J.~C.,  {Gil de Paz} A.,  {Boissier} S.,  {Zamorano} J.,  {Jarrett} T.,  {Gallego} J.,   {Madore} B.~F.,  2007a, \mn@doi [\apj] {10.1086/511812}, \href {https://ui.adsabs.harvard.edu/abs/2007ApJ...658.1006M} {658, 1006}

\bibitem[\protect\citeauthoryear{{Mu{\~n}oz-Mateos}, {Gil de Paz}, {Boissier}, {Zamorano}, {Jarrett}, {Gallego}  \& {Madore}}{{Mu{\~n}oz-Mateos} et~al.}{2007b}]{Mateos2007}
{Mu{\~n}oz-Mateos} J.~C.,  {Gil de Paz} A.,  {Boissier} S.,  {Zamorano} J.,  {Jarrett} T.,  {Gallego} J.,   {Madore} B.~F.,  2007b, \mn@doi [\apj] {10.1086/511812}, \href {https://ui.adsabs.harvard.edu/abs/2007ApJ...658.1006M} {658, 1006}

\bibitem[\protect\citeauthoryear{{Pak}, {Lee}, {Jeong}, {Kim}, {Smith}  \& {Lee}}{{Pak} et~al.}{2019}]{Pak2019}
{Pak} M.,  {Lee} J.~H.,  {Jeong} H.,  {Kim} S.,  {Smith} R.,   {Lee} H.-R.,  2019, \mn@doi [\apj] {10.3847/1538-4357/ab2ad6}, \href {https://ui.adsabs.harvard.edu/abs/2019ApJ...880..149P} {880, 149}

\bibitem[\protect\citeauthoryear{{Poggianti}, {Smail}, {Dressler}, {Couch}, {Barger}, {Butcher}, {Ellis}  \& {Oemler}}{{Poggianti} et~al.}{1999}]{Poggianti1999}
{Poggianti} B.~M.,  {Smail} I.,  {Dressler} A.,  {Couch} W.~J.,  {Barger} A.~J.,  {Butcher} H.,  {Ellis} R.~S.,   {Oemler} Augustus J.,  1999, \mn@doi [\apj] {10.1086/307322}, \href {https://ui.adsabs.harvard.edu/abs/1999ApJ...518..576P} {518, 576}

\bibitem[\protect\citeauthoryear{{Querejeta}, {Eliche-Moral}, {Tapia}, {Borlaff}, {Rodr{\'\i}guez-P{\'e}rez}, {Zamorano}  \& {Gallego}}{{Querejeta} et~al.}{2015}]{Querejeta2015}
{Querejeta} M.,  {Eliche-Moral} M.~C.,  {Tapia} T.,  {Borlaff} A.,  {Rodr{\'\i}guez-P{\'e}rez} C.,  {Zamorano} J.,   {Gallego} J.,  2015, \mn@doi [\aap] {10.1051/0004-6361/201424303}, \href {https://ui.adsabs.harvard.edu/abs/2015A&A...573A..78Q} {573, A78}

\bibitem[\protect\citeauthoryear{{Quilis}, {Moore}  \& {Bower}}{{Quilis} et~al.}{2000}]{Quilis2000}
{Quilis} V.,  {Moore} B.,   {Bower} R.,  2000, \mn@doi [Science] {10.1126/science.288.5471.1617}, \href {https://ui.adsabs.harvard.edu/abs/2000Sci...288.1617Q} {288, 1617}

\bibitem[\protect\citeauthoryear{{Rathore}, {Kumar}, {Mishra}, {Wadadekar}  \& {Bait}}{{Rathore} et~al.}{2022}]{Rathore2022}
{Rathore} H.,  {Kumar} K.,  {Mishra} P.~K.,  {Wadadekar} Y.,   {Bait} O.,  2022, \mn@doi [\mnras] {10.1093/mnras/stac871}, \href {https://ui.adsabs.harvard.edu/abs/2022MNRAS.513..389R} {513, 389}

\bibitem[\protect\citeauthoryear{{Robaina}, {Hoyle}, {Gallazzi}, {Jim{\'e}nez}, {van der Wel}  \& {Verde}}{{Robaina} et~al.}{2012}]{Robaina2012}
{Robaina} A.~R.,  {Hoyle} B.,  {Gallazzi} A.,  {Jim{\'e}nez} R.,  {van der Wel} A.,   {Verde} L.,  2012, \mn@doi [\mnras] {10.1111/j.1365-2966.2012.21804.x}, \href {https://ui.adsabs.harvard.edu/abs/2012MNRAS.427.3006R} {427, 3006}

\bibitem[\protect\citeauthoryear{{Robertson}, {Bullock}, {Cox}, {Di Matteo}, {Hernquist}, {Springel}  \& {Yoshida}}{{Robertson} et~al.}{2006}]{Robertson2006}
{Robertson} B.,  {Bullock} J.~S.,  {Cox} T.~J.,  {Di Matteo} T.,  {Hernquist} L.,  {Springel} V.,   {Yoshida} N.,  2006, \mn@doi [\apj] {10.1086/504412}, \href {https://ui.adsabs.harvard.edu/abs/2006ApJ...645..986R} {645, 986}

\bibitem[\protect\citeauthoryear{{Roca-F{\`a}brega}, {Valenzuela}, {Figueras}, {Romero-G{\'o}mez}, {Vel{\'a}zquez}, {Antoja}  \& {Pichardo}}{{Roca-F{\`a}brega} et~al.}{2013}]{Roca2013}
{Roca-F{\`a}brega} S.,  {Valenzuela} O.,  {Figueras} F.,  {Romero-G{\'o}mez} M.,  {Vel{\'a}zquez} H.,  {Antoja} T.,   {Pichardo} B.,  2013, \mn@doi [\mnras] {10.1093/mnras/stt643}, \href {https://ui.adsabs.harvard.edu/abs/2013MNRAS.432.2878R} {432, 2878}

\bibitem[\protect\citeauthoryear{{S{\'a}nchez} et~al.,}{{S{\'a}nchez} et~al.}{2016}]{Sanchez2016}
{S{\'a}nchez} S.~F.,  et~al., 2016, \mn@doi [\rmxaa] {10.48550/arXiv.1509.08552}, \href {https://ui.adsabs.harvard.edu/abs/2016RMxAA..52...21S} {52, 21}

\bibitem[\protect\citeauthoryear{{Schawinski} et~al.,}{{Schawinski} et~al.}{2014}]{Schawinski2014}
{Schawinski} K.,  et~al., 2014, \mn@doi [\mnras] {10.1093/mnras/stu327}, \href {https://ui.adsabs.harvard.edu/abs/2014MNRAS.440..889S} {440, 889}

\bibitem[\protect\citeauthoryear{{Sellwood}}{{Sellwood}}{2011}]{Sellwood2011}
{Sellwood} J.~A.,  2011, \mn@doi [\mnras] {10.1111/j.1365-2966.2010.17545.x}, \href {https://ui.adsabs.harvard.edu/abs/2011MNRAS.410.1637S} {410, 1637}

\bibitem[\protect\citeauthoryear{{Sellwood}}{{Sellwood}}{2013}]{Sellwood2013}
{Sellwood} J.~A.,  2013, in {Oswalt} T.~D.,  {Gilmore} G.,  eds, , Vol.~5, Planets, Stars and Stellar Systems. Volume 5: Galactic Structure and Stellar Populations.
p.~923, \mn@doi{10.1007/978-94-007-5612-0_18}

\bibitem[\protect\citeauthoryear{{Sellwood} \& {Carlberg}}{{Sellwood} \& {Carlberg}}{1984}]{Sellwood1984}
{Sellwood} J.~A.,  {Carlberg} R.~G.,  1984, \mn@doi [\apj] {10.1086/162176}, \href {https://ui.adsabs.harvard.edu/abs/1984ApJ...282...61S} {282, 61}

\bibitem[\protect\citeauthoryear{{Shen}, {Mo}, {White}, {Blanton}, {Kauffmann}, {Voges}, {Brinkmann}  \& {Csabai}}{{Shen} et~al.}{2003}]{Shen2003}
{Shen} S.,  {Mo} H.~J.,  {White} S. D.~M.,  {Blanton} M.~R.,  {Kauffmann} G.,  {Voges} W.,  {Brinkmann} J.,   {Csabai} I.,  2003, \mn@doi [\mnras] {10.1046/j.1365-8711.2003.06740.x}, \href {https://ui.adsabs.harvard.edu/abs/2003MNRAS.343..978S} {343, 978}

\bibitem[\protect\citeauthoryear{{Shimakawa}, {Tanaka}, {Bottrell}, {Wu}, {Chang}, {Toba}  \& {Ali}}{{Shimakawa} et~al.}{2022}]{Shimakawa2022}
{Shimakawa} R.,  {Tanaka} M.,  {Bottrell} C.,  {Wu} P.-F.,  {Chang} Y.-Y.,  {Toba} Y.,   {Ali} S.,  2022, \mn@doi [\pasj] {10.1093/pasj/psac023}, \href {https://ui.adsabs.harvard.edu/abs/2022PASJ...74..612S} {74, 612}

\bibitem[\protect\citeauthoryear{{Shlosman}, {Peletier}  \& {Knapen}}{{Shlosman} et~al.}{2000}]{Shlosman2000ApJ}
{Shlosman} I.,  {Peletier} R.~F.,   {Knapen} J.~H.,  2000, \mn@doi [\apjl] {10.1086/312716}, \href {https://ui.adsabs.harvard.edu/abs/2000ApJ...535L..83S} {535, L83}

\bibitem[\protect\citeauthoryear{{Sparre} \& {Springel}}{{Sparre} \& {Springel}}{2017}]{Sparre2017}
{Sparre} M.,  {Springel} V.,  2017, \mn@doi [\mnras] {10.1093/mnras/stx1516}, \href {https://ui.adsabs.harvard.edu/abs/2017MNRAS.470.3946S} {470, 3946}

\bibitem[\protect\citeauthoryear{{Speagle}, {Steinhardt}, {Capak}  \& {Silverman}}{{Speagle} et~al.}{2014}]{Speagle2014ApJS}
{Speagle} J.~S.,  {Steinhardt} C.~L.,  {Capak} P.~L.,   {Silverman} J.~D.,  2014, \mn@doi [\apjs] {10.1088/0067-0049/214/2/15}, \href {https://ui.adsabs.harvard.edu/abs/2014ApJS..214...15S} {214, 15}

\bibitem[\protect\citeauthoryear{{Springel} \& {Hernquist}}{{Springel} \& {Hernquist}}{2005}]{Springel2005}
{Springel} V.,  {Hernquist} L.,  2005, \mn@doi [\apjl] {10.1086/429486}, \href {https://ui.adsabs.harvard.edu/abs/2005ApJ...622L...9S} {622, L9}

\bibitem[\protect\citeauthoryear{{Stark} et~al.,}{{Stark} et~al.}{2021}]{Stark2021}
{Stark} D.~V.,  et~al., 2021, \mn@doi [\mnras] {10.1093/mnras/stab566}, \href {https://ui.adsabs.harvard.edu/abs/2021MNRAS.503.1345S} {503, 1345}

\bibitem[\protect\citeauthoryear{{Strateva} et~al.,}{{Strateva} et~al.}{2001}]{Strateva2001}
{Strateva} I.,  et~al., 2001, \mn@doi [\aj] {10.1086/323301}, \href {https://ui.adsabs.harvard.edu/abs/2001AJ....122.1861S} {122, 1861}

\bibitem[\protect\citeauthoryear{{Tapia}, {Eliche-Moral}, {Aceves}, {Rodr{\'\i}guez-P{\'e}rez}, {Borlaff}  \& {Querejeta}}{{Tapia} et~al.}{2017}]{Tapia2017}
{Tapia} T.,  {Eliche-Moral} M.~C.,  {Aceves} H.,  {Rodr{\'\i}guez-P{\'e}rez} C.,  {Borlaff} A.,   {Querejeta} M.,  2017, \mn@doi [\aap] {10.1051/0004-6361/201628821}, \href {https://ui.adsabs.harvard.edu/abs/2017A&A...604A.105T} {604, A105}

\bibitem[\protect\citeauthoryear{{Thanjavur}, {Simard}, {Bluck}  \& {Mendel}}{{Thanjavur} et~al.}{2016}]{KT2016}
{Thanjavur} K.,  {Simard} L.,  {Bluck} A. F.~L.,   {Mendel} T.,  2016, \mn@doi [\mnras] {10.1093/mnras/stw495}, \href {https://ui.adsabs.harvard.edu/abs/2016MNRAS.459...44T} {459, 44}

\bibitem[\protect\citeauthoryear{{Thomas}, {Maraston}, {Bender}  \& {Mendes de Oliveira}}{{Thomas} et~al.}{2005}]{Thomas2005}
{Thomas} D.,  {Maraston} C.,  {Bender} R.,   {Mendes de Oliveira} C.,  2005, \mn@doi [\apj] {10.1086/426932}, \href {https://ui.adsabs.harvard.edu/abs/2005ApJ...621..673T} {621, 673}

\bibitem[\protect\citeauthoryear{{Tojeiro} et~al.,}{{Tojeiro} et~al.}{2013}]{Tojeiro2013}
{Tojeiro} R.,  et~al., 2013, \mn@doi [\mnras] {10.1093/mnras/stt484}, \href {https://ui.adsabs.harvard.edu/abs/2013MNRAS.432..359T} {432, 359}

\bibitem[\protect\citeauthoryear{{Tortora}, {Antonuccio-Delogu}, {Kaviraj}, {Silk}, {Romeo}  \& {Becciani}}{{Tortora} et~al.}{2009}]{Tortora2009}
{Tortora} C.,  {Antonuccio-Delogu} V.,  {Kaviraj} S.,  {Silk} J.,  {Romeo} A.~D.,   {Becciani} U.,  2009, \mn@doi [\mnras] {10.1111/j.1365-2966.2009.14718.x}, \href {https://ui.adsabs.harvard.edu/abs/2009MNRAS.396...61T} {396, 61}

\bibitem[\protect\citeauthoryear{{Tortora}, {Napolitano}, {Cardone}, {Capaccioli}, {Jetzer}  \& {Molinaro}}{{Tortora} et~al.}{2010}]{Tortora2010}
{Tortora} C.,  {Napolitano} N.~R.,  {Cardone} V.~F.,  {Capaccioli} M.,  {Jetzer} P.,   {Molinaro} R.,  2010, \mn@doi [\mnras] {10.1111/j.1365-2966.2010.16938.x}, \href {https://ui.adsabs.harvard.edu/abs/2010MNRAS.407..144T} {407, 144}

\bibitem[\protect\citeauthoryear{{Trager}, {Worthey}, {Faber}, {Burstein}  \& {Gonz{\'a}lez}}{{Trager} et~al.}{1998}]{Trager1998}
{Trager} S.~C.,  {Worthey} G.,  {Faber} S.~M.,  {Burstein} D.,   {Gonz{\'a}lez} J.~J.,  1998, \mn@doi [\apjs] {10.1086/313099}, \href {https://ui.adsabs.harvard.edu/abs/1998ApJS..116....1T} {116, 1}

\bibitem[\protect\citeauthoryear{{Trussler}, {Maiolino}, {Maraston}, {Peng}, {Thomas}, {Goddard}  \& {Lian}}{{Trussler} et~al.}{2020}]{Trussler2020}
{Trussler} J.,  {Maiolino} R.,  {Maraston} C.,  {Peng} Y.,  {Thomas} D.,  {Goddard} D.,   {Lian} J.,  2020, \mn@doi [\mnras] {10.1093/mnras/stz3286}, \href {https://ui.adsabs.harvard.edu/abs/2020MNRAS.491.5406T} {491, 5406}

\bibitem[\protect\citeauthoryear{{Tully} \& {Pierce}}{{Tully} \& {Pierce}}{2000}]{Tully2000}
{Tully} R.~B.,  {Pierce} M.~J.,  2000, \mn@doi [\apj] {10.1086/308700}, \href {https://ui.adsabs.harvard.edu/abs/2000ApJ...533..744T} {533, 744}

\bibitem[\protect\citeauthoryear{{Vazdekis}, {Cenarro}, {Gorgas}, {Cardiel}  \& {Peletier}}{{Vazdekis} et~al.}{2003}]{Vazdekis2003}
{Vazdekis} A.,  {Cenarro} A.~J.,  {Gorgas} J.,  {Cardiel} N.,   {Peletier} R.~F.,  2003, \mn@doi [\mnras] {10.1046/j.1365-8711.2003.06413.x}, \href {https://ui.adsabs.harvard.edu/abs/2003MNRAS.340.1317V} {340, 1317}

\bibitem[\protect\citeauthoryear{{Vikram}, {Wadadekar}, {Kembhavi}  \& {Vijayagovindan}}{{Vikram} et~al.}{2010}]{Vikram2010}
{Vikram} V.,  {Wadadekar} Y.,  {Kembhavi} A.~K.,   {Vijayagovindan} G.~V.,  2010, \mn@doi [\mnras] {10.1111/j.1365-2966.2010.17426.x}, \href {https://ui.adsabs.harvard.edu/abs/2010MNRAS.409.1379V} {409, 1379}

\bibitem[\protect\citeauthoryear{{Wang} et~al.,}{{Wang} et~al.}{2016}]{Wang2016}
{Wang} H.,  et~al., 2016, \mn@doi [\apj] {10.3847/0004-637X/831/2/164}, \href {https://ui.adsabs.harvard.edu/abs/2016ApJ...831..164W} {831, 164}

\bibitem[\protect\citeauthoryear{{Wang} et~al.,}{{Wang} et~al.}{2022}]{Wang2022}
{Wang} L.,  et~al., 2022, \mn@doi [\mnras] {10.1093/mnras/stac2292}, \href {https://ui.adsabs.harvard.edu/abs/2022MNRAS.516.2337W} {516, 2337}

\bibitem[\protect\citeauthoryear{{Westfall} et~al.,}{{Westfall} et~al.}{2019}]{Westfall2019}
{Westfall} K.~B.,  et~al., 2019, \mn@doi [\aj] {10.3847/1538-3881/ab44a2}, \href {https://ui.adsabs.harvard.edu/abs/2019AJ....158..231W} {158, 231}

\bibitem[\protect\citeauthoryear{{Whitaker} et~al.,}{{Whitaker} et~al.}{2017}]{Whitaker2017ApJ}
{Whitaker} K.~E.,  et~al., 2017, \mn@doi [\apj] {10.3847/1538-4357/aa6258}, \href {https://ui.adsabs.harvard.edu/abs/2017ApJ...838...19W} {838, 19}

\bibitem[\protect\citeauthoryear{{White}}{{White}}{1980}]{White1980}
{White} S.~D.~M.,  1980, \mn@doi [\mnras] {10.1093/mnras/191.1.1P}, \href {https://ui.adsabs.harvard.edu/abs/1980MNRAS.191P...1W} {191, 1P}

\bibitem[\protect\citeauthoryear{{Worthey} \& {Ottaviani}}{{Worthey} \& {Ottaviani}}{1997}]{Worthey1997}
{Worthey} G.,  {Ottaviani} D.~L.,  1997, \mn@doi [\apjs] {10.1086/313021}, \href {https://ui.adsabs.harvard.edu/abs/1997ApJS..111..377W} {111, 377}

\bibitem[\protect\citeauthoryear{{Wright} et~al.,}{{Wright} et~al.}{2010}]{Wright2010}
{Wright} E.~L.,  et~al., 2010, \mn@doi [\aj] {10.1088/0004-6256/140/6/1868}, \href {https://ui.adsabs.harvard.edu/abs/2010AJ....140.1868W} {140, 1868}

\bibitem[\protect\citeauthoryear{{Xu}, {Gu}, {Lu}, {Ge}, {Xiao}  \& {Contini}}{{Xu} et~al.}{2022}]{Xu2022}
{Xu} K.,  {Gu} Q.,  {Lu} S.,  {Ge} X.,  {Xiao} M.,   {Contini} E.,  2022, \mn@doi [\mnras] {10.1093/mnras/stab3013}, \href {https://ui.adsabs.harvard.edu/abs/2022MNRAS.509.1237X} {509, 1237}

\bibitem[\protect\citeauthoryear{{Yang}, {Mo}, {van den Bosch}, {Pasquali}, {Li}  \& {Barden}}{{Yang} et~al.}{2007}]{Yang2007}
{Yang} X.,  {Mo} H.~J.,  {van den Bosch} F.~C.,  {Pasquali} A.,  {Li} C.,   {Barden} M.,  2007, \mn@doi [\apj] {10.1086/522027}, \href {https://ui.adsabs.harvard.edu/abs/2007ApJ...671..153Y} {671, 153}

\bibitem[\protect\citeauthoryear{{York} et~al.,}{{York} et~al.}{2000}]{York2000}
{York} D.~G.,  et~al., 2000, \mn@doi [\aj] {10.1086/301513}, \href {https://ui.adsabs.harvard.edu/abs/2000AJ....120.1579Y} {120, 1579}

\bibitem[\protect\citeauthoryear{{Zhou}, {Li}, {Hao}, {Guo}, {Mo}  \& {Xia}}{{Zhou} et~al.}{2021}]{Zhou2021}
{Zhou} S.,  {Li} C.,  {Hao} C.-N.,  {Guo} R.,  {Mo} H.,   {Xia} X.,  2021, \mn@doi [\apj] {10.3847/1538-4357/ac06cc}, \href {https://ui.adsabs.harvard.edu/abs/2021ApJ...916...38Z} {916, 38}

\bibitem[\protect\citeauthoryear{{Zolotov} et~al.,}{{Zolotov} et~al.}{2015}]{Zolotov2015}
{Zolotov} A.,  et~al., 2015, \mn@doi [\mnras] {10.1093/mnras/stv740}, \href {https://ui.adsabs.harvard.edu/abs/2015MNRAS.450.2327Z} {450, 2327}

\bibitem[\protect\citeauthoryear{{van der Wel} et~al.,}{{van der Wel} et~al.}{2014}]{VanderWel2014}
{van der Wel} A.,  et~al., 2014, \mn@doi [\apj] {10.1088/0004-637X/788/1/28}, \href {https://ui.adsabs.harvard.edu/abs/2014ApJ...788...28V} {788, 28}

\makeatother
\end{thebibliography}




\bsp	
\label{lastpage}
\end{CJK*}
\end{document}